\documentclass[11pt,a4paper,oneside,openright]{book}

\usepackage{mathptmx}
\DeclareMathAlphabet{\altmathcal}{OMS}{cmsy}{m}{n}
\usepackage[margin=2.5cm]{geometry}

\usepackage{setspace}
\usepackage{fancyhdr}
\fancypagestyle{plain}{
  \fancyhead{}
  
}

\usepackage{hyperref}
\usepackage{tensor}
\usepackage{scalerel,amssymb}

\usepackage{graphicx}

\usepackage[nottoc]{tocbibind}

\usepackage{indentfirst}

\usepackage{enumitem}

\usepackage[font=footnotesize]{caption}

\usepackage[toc,page]{appendix}

\usepackage{pdfpages}

\usepackage{fmtcount}

\usepackage{url}

\usepackage{amsmath}
\usepackage{bm}

\usepackage[english]{babel}

\usepackage[utf8]{inputenc}

\usepackage{filecontents}
\usepackage{float}

\usepackage{subcaption}
\usepackage{eurosym}

\usepackage{booktabs}

\usepackage[T1]{fontenc}

\makeatletter
\providecommand{\keywords}[1]{\textbf{Keywords:} #1}
\providecommand{\keywordsP}[1]{\textbf{Palavras chave:} #1}

\begin{document}
%\onehalfspacing
\begin{sloppy}

%    \makeatletter
%    \renewcommand{\@makechapterhead}[1]{%
%    \vspace*{50 pt}%
%    {\setlength{\parindent}{0pt} \raggedright \normalfont
%    \bfseries\Huge
%    \ifnum \value{secnumdepth}>1 
%        \if@mainmatter\thechapter.\ \fi%
%    \fi
%    #1\par\nobreak\vspace{40 pt}}}
%    \makeatother

\frontmatter
%-----------------------------------------------------------------
%
% Capa (FRENTE)
%
%------------------------------------------------------------------

%Esta capa é apenas um exemplo. Devem retirar a capa oficial do site da faculdade, alterar de acordo com o trabalho realizado, guardar como pdf e fazer o upload para o overleaf. Deve ser depois substituído no comando seguinte (includepdf) pelo seu respectivo nome.

\begin{titlepage}
\includepdf[]{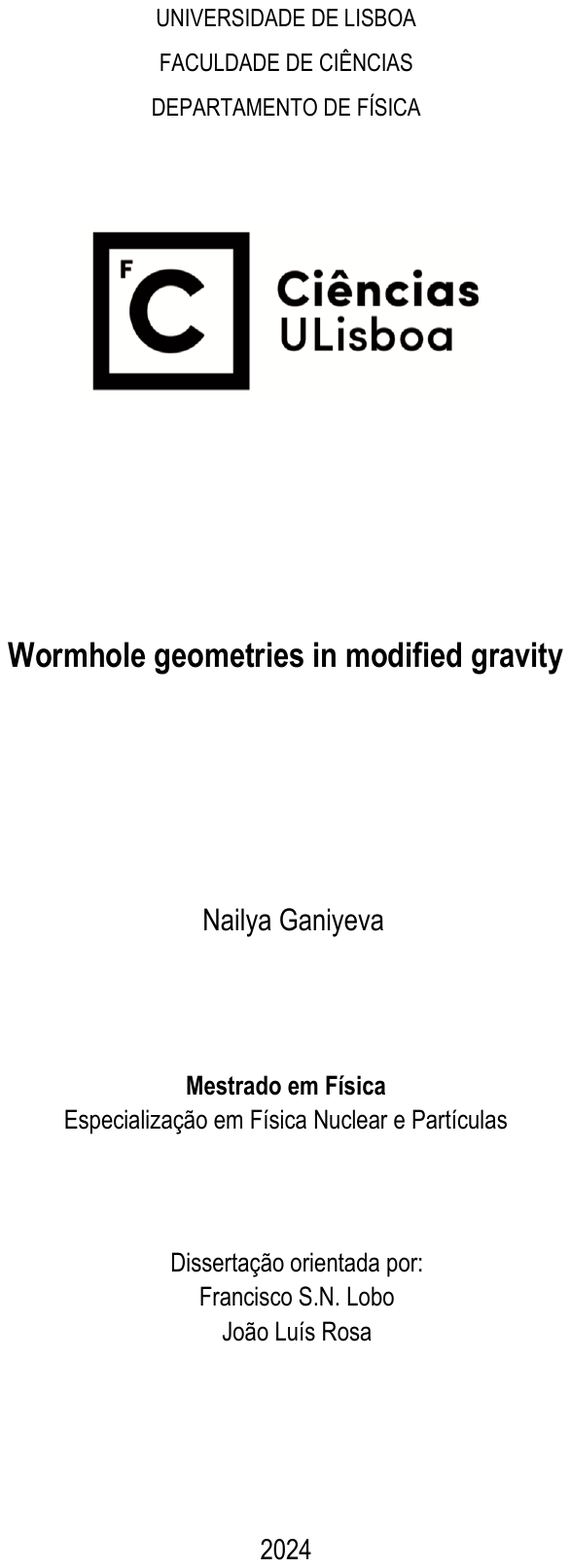}
\end{titlepage}

%-----------------------------------------------------------------
%
% Capa (TRÁS)
%
%------------------------------------------------------------------
\clearpage \thispagestyle{empty}\mbox{}\clearpage
%\pagenumbering{roman}
%\newpage
%\thispagestyle{plain}
%\mbox{}

%-----------------------------------------------------------------
%
% RESUMO
%
%------------------------------------------------------------------
\newpage
\thispagestyle{plain}
\chapter*{Resumo}

Nesta tese, concentramo-nos no estudo de wormholes, que são estruturas teóricas que conectam duas regiões distintas do espaço-tempo através de uma estrutura geométrica conhecida como garganta. Dado a ausência de evidências observacionais destes objetos, é necessário realizar esta análise dentro da melhor teoria que temos de momento para descrever a gravidade. Assim, recorremos à Relatividade Geral (RG), formulada por Einstein em 1915. Explorando então esta abordagem teórica, a análise revela que a matéria necessária para suportar estas geometrias é caracterizada como matéria exótica, designada desta forma por violar a condição de energia nula. Na verdade, dado que a condição de energia nula é a condição de energia mais fraca, a sua violação implica a violação das outras condições de energia, nomeadamente a condição de energia fraca, forte e dominante. Contudo, uma vez que esta matéria exótica não foi observada em sistemas astrofísicos, implica que as soluções de wormholes obtidas no contexto da teoria da RG carecem sentido físico.

Face a este problema, surge então a necessidade de procurar soluções que minimizem ou, idealmente, evitem o uso de matéria exótica. Uma possível via para o resolver passa por considerar teorias de gravidade modificadas, que introduzem termos gravitacionais adicionais relativamente à RG, de tal forma que a ação modificada consiga simultaneamente dar origem à geometria do wormhole e ser sustentada por matéria que obedece à condição de energia nula. Portanto, o objetivo desta tese é obter soluções de wormholes considerando uma teoria de gravidade modificada, em particular a teoria de gravidade $f(R,\mathcal{T})$, sendo $R$ o escalar de Ricci e $\mathcal{T}=T_{ab}T^{ab}$, onde $T_{ab}$ é o tensor de energia-momento. 

Além disso, sabemos hoje em dia que a RG não será a teoria final que descreve a gravidade, uma vez que apresenta limitações como a necessidade de introduzir matéria escura e energia escura, para descrever fenómenos como a rotação das galáxias e a aceleração do Universo, caso contrário não concorda com os dados observacionais. Estas evidências experimentais levam mais uma vez a considerar extensões ou modificações da teoria. Neste sentido, o nosso estudo não só visa encontrar soluções de wormholes no contexto da teoria de gravidade $f(R,\mathcal{T})$, como também avaliar se esta teoria em particular se justifica como uma modificação válida e fisicamente significativa da Relatividade Geral.

Para obtermos soluções fisicamente plausíveis que descrevam objetos localizados, poderá ser necessário recorrer às condições de junção, obtidas pelo formalismo distribucional. Dado que estas condições de junção são dependentes da teoria, o nosso segundo objetivo deste trabalho consiste em derivá-las especificamente para o caso da teoria de gravidade $f(R,\mathcal{T})$. Conforme apresentado no capítulo \ref{ch:results}, as soluções obtidas não se encontram localizadas. Observa-se que, apesar das componentes da métrica exibirem comportamento assintoticamente plano, elas não tendem para o vácuo assintoticamente. Portanto, necessitamos das condições de junção para unir o espaço-tempo associado à região do wormhole a um espaço-tempo externo associado ao vácuo, num determinado raio finito. Desta forma, obtemos finalmente, soluções de wormholes localizadas que obedecem a todas as condições de energia. 

No capítulo \ref{ch:intro}, inicia-se este estudo revendo e reforçando a sua motivação, seguindo depois para uma breve revisão histórica da literatura sobre wormholes. Começamos com as primeiras considerações desenvolvidas por Einstein e Rosen, posteriormente por Wheeler, até chegar ao trabalho pioneiro desenvolvido por Morris e Thorne, onde obtiveram pela primeira vez soluções de wormholes atravessáveis e analisaram as suas consequências, partindo de uma estratégia diferente na qual consideram primeiro a geometria do wormhole e depois obtêm as componentes da matéria através das equações de campo. A revisão continua abrangendo agora estudos mais modernos, nos quais esses objetos são analisados no contexto de diferentes teorias de gravidade modificadas. Por fim, terminamos o capítulo destacando a importância do nosso trabalho e como contribui para o desenvolvimento desta área.

No capítulo \ref{ch:framework}, estruturamos a discussão em duas secções. Começamos por uma introdução à teoria da Relatividade Geral, seguida por uma aplicação prática, centrada no estudo de wormholes no contexto desta teoria. Na primeira parte, começamos por abordar os conceitos fundamentais da Relatividade Geral, introduzindo quantidades tensoriais geométricas e considerações necessárias para a dedução das equações de Einstein. Em seguida, repetimos o processo de dedução das Equações de Einstein desta vez utilizando o formalismo Lagrangiano. Esta abordagem será vantajosa no capítulo \ref{ch: MGT}, já que a modificação de uma teoria de gravidade é mais simples considerando este formalismo. Por último, concluímos esta primeira parte introduzindo o formalismo distribucional, necessário para derivar as condições de junção em GR. Na segunda parte do capítulo, apresentamos a métrica que caracteriza a geometria dos wormholes e derivamos as suas equações de campo correspondentes. Ao analisarmos as componentes da matéria que sustenta esta geometria, devemos verificar se estas obedecem às condições de energia. Contudo, como já foi referido anteriormente, concluímos que a condição de energia nula é violada, indicando a presença de matéria exótica.

No capítulo \ref{ch: MGT}, organizamos a discussão em três secções. Na primeira secção, apresentamos a teoria de gravidade modificada $f(R,\mathcal{T})$, derivando as suas equações de campo modificadas e a equação de conservação para o tensor de energia-momento $T_{ab}$. Em seguida, procedemos à derivação das equações de campo relativamente ao wormhole, considerando uma forma linear tanto em $R$ quanto em $\mathcal{T}$ para a função $f(R,\mathcal{T})$, isto é, assumimos que $f(R,\mathcal{T})=R+\gamma \mathcal{T}$, onde $\gamma$ é uma constante de acoplamento. Dada a complexidade destas equações, recorremos a um algoritmo recursivo para a sua resolução, dado seguinte forma:  Começamos por escolher os valores das constantes e resolvemos as equações de campo para obter os valores das componentes de matéria num determinado raio inicial, denotado por $r_0$. Para cada solução obtida, incrementamos o raio para $r_{n+1}=r_n+\epsilon$, onde $\epsilon$ é um valor pequeno, e resolvemos as equações de campo para o novo raio $r_{n+1}$. Repetimos este processo até atingir um raio suficientemente grande, obtendo assim soluções analíticas para as componentes de matéria. Na segunda secção, derivamos as condições de junção para este caso linear. Dado que as soluções obtidas não apresentam comportamento assintoticamente nulo nas componentes da matéria, como apresentado no capítulo \ref{ch:results}, o que indica que o objeto não se encontra localizado, realizamos a junção deste espaço-tempo interior associado ao wormhole com um espaço-tempo exterior correspondente a um vácuo. Finalmente, na terceira secção, generalizamos esta abordagem para funções $f(R,\mathcal{T})$ que incluem desta vez termos $\mathcal{T}$ de grau superior. Derivamos assim as soluções de wormhole para este caso, juntamente com as respetivas condições de junção, que se verificam ser idênticas às do caso linear, contanto que não haja termos cruzados entre $R$ e $\mathcal{T}$.

No capítulo \ref{ch:results}, apresentamos exemplos de soluções obtidas, bem como os resultados provenientes da aplicação das condições de junção. Começamos pelo caso da função linear $f(R,\mathcal{T})= R+\gamma \mathcal{T}$ e, em seguida, procedemos para o caso quadrático $f(R,\mathcal{T})= R+ \sigma \mathcal{T}^2$. Em ambos os casos, observamos que as componentes da matéria, embora assintoticamente planas, não atingem o vácuo assintoticamente. Assim, como já foi referido, é necessário uma junção com o espaço-tempo exterior correspondente ao vácuo. Dado que, como concluímos no capítulo \ref{ch: MGT}, as condições de junção são as mesmas, os resultados da união são os mesmos para ambos os casos. Além disso, nesta seção, destacamos que, no caso das funções $f(R,\mathcal{T})= R+\gamma \mathcal{T}$, a constante de acoplamento $\gamma$ deve ser necessariamente negativa para obtermos soluções que obedecem a todas as condições de energia. No entanto, no caso de funções $f(R,\mathcal{T})= R+ \gamma \mathcal{T} + \sigma \mathcal{T}^n$, a constante $\gamma$ pode assumir valores nulos ou positivos, desde que $\sigma$ seja negativo. Em resumo, conseguimos obter soluções localizadas de wormholes que obedecem a todas as condições de energia.

\keywordsP{Wormholes, teorias de gravidade modificadas, condições de junção.}

%-----------------------------------------------------------------
%
% ABSTRACT
%
%------------------------------------------------------------------
\newpage
\thispagestyle{plain}
\chapter*{Abstract}

In this thesis, we investigate traversable wormhole spacetimes within the context of a covariant generalization of Einstein's General Relativity, namely the energy-momentum squared gravity, denoted as $f\left(R,\mathcal{T}\right)$. Here, $R$ represents the Ricci scalar and $\mathcal{T}=T_{ab}T^{ab}$, where $T_{ab}$ is the energy-momentum tensor. Specifically considering the linear form $f\left(R,\mathcal{T}\right)=R+\gamma \mathcal{T}$, we demonstrate the existence of numerous wormhole solutions, wherein the matter fields satisfy all energy conditions, these being the null, weak, strong, and dominant energy conditions. Remarkably, these solutions do not require fine-tuning of the free parameters inherent to the model. Given the complicated nature of the field equations, we develop an analytical recursive algorithm to derive these solutions. One limitation is that these solutions lack natural localization, necessitating then a matching with an external vacuum spacetime. To address this, we derive the junction conditions for the theory, establishing that any matching between two spacetimes must be smooth, i.e. without the presence of thin-shells at the boundary. Ultimately, applying these junction conditions allows us to match the interior wormhole spacetime with an exterior vacuum described by the Schwarzschild solution. Therefore, we obtain traversable, localized, static, and spherically symmetric wormhole solutions that satisfy all energy conditions across the entire spacetime range. Additionally, we demonstrate that the approach used in this study can be easily generalized to more complicated dependencies of the action on $\mathcal T$, provided there are no crossed terms between $R$ and $\mathcal{T}$.

\keywords{Wormholes, modified theories of gravity, junction conditions.} 
%-----------------------------------------------------------------
%
% AGRADECIMENTOS
%
%------------------------------------------------------------------
\newpage
\thispagestyle{plain}
\chapter*{Acknowledgements}

First and foremost, my deepest gratitude goes to my supervisors. Francisco Lobo, for embracing this challenge with me, for the engaging discussions, always sharing your profound knowledge and boundless curiosity (often leaving me with more questions than answers), and for maintaining an infectious enthusiasm till the very end of this work. And, Jo\~{a}o Lu\'{i}s, for your guidance and constant support, infinite patience, invaluable teachings throughout this thesis and always finding time for my shenanigans. Both of you are inspirations, not only as physicists, but as human beings. 
\paragraph{}
Secondly, I would like to thank my family. Due to a language barrier, mostly caused by me, my father believes that I am trying to be an astronaut, and my mother that I am finishing Einstein's work. Not sure if they are joking, but nevertheless, their blind support and trust in every step means the whole world. And my friends, who, in the midst of laughs and cries, always find the words to push me to be the best version of myself. 
\paragraph{}
Finally, to the person that made this thesis possible and paradoxically was my biggest distraction, Leo G\"{u}ndersen. Thank you for sharing this journey with me, and I can't wait to share many more. 

\paragraph{}
%---------------------------------------------------------------------------------------------------
%
% ÍNDICE
%
%------------------------------------------------------------------
% \newpage
\thispagestyle{plain}
\renewcommand{\contentsname}{Table of Contents}
\tableofcontents
\newpage
\thispagestyle{plain}
\listoffigures
%\newpage
%\thispagestyle{plain}
%\listoftables
%\clearpage \thispagestyle{plain}\mbox{}\clearpage
\newpage
\thispagestyle{plain}
\chapter{Preface}

The official research presented in this Master's thesis has been carried out at the Instituto de Astrof\'{i}sica e Ci\^{e}ncias do Espaço (IA) and in the Physics Department of Faculdade de Ci\^{e}ncias da Universidade de Lisboa, under the projects UIDB/04434/2020 and UIDP/04434/2020.
\paragraph{}
This work was conducted in collaboration with my supervisor, Professor Francisco S.N. Lobo, and my co-supervisor, Doctor Jo\~{a}o Lu\'{i}s Rosa, from which Chapters \ref{ch:modifiedgravity} and \ref{ch:results} have been published in The European Physical Journal C, namely:
\paragraph{}
- J.~L.~Rosa, N.~Ganiyeva and F.~S.~N.~Lobo,
``Non-exotic traversable wormholes in $f\left( R,T_{ab}T^{ab}\right) $ gravity,''
Eur. Phys. J. C \textbf{83} (2023) no.11, 1040
%doi:10.1140/epjc/s10052-023-12232-0
[arXiv:2309.08768 [gr-qc]].
\paragraph{}
Later presented in the following conference:
\paragraph{}
- N.~Ganiyeva, ``Wormhole geometries in modified gravity'', 3 DAY Mission: SP4C3 Exploration by Physis, October 15th, 2023, Faculdade de Ci\^{e}ncias da Universidade de Lisboa, Lisbon, Portugal.

\mainmatter
\pagenumbering{arabic} %numeros em numeração árabe
%-----------------------------------------------------------------
%
% CHAPTER 1: INTRODUCTION
%
%------------------------------------------------------------------
\newpage
\thispagestyle{plain}

%%%%%%%%%%%%%%%%%%%%%%%%%%%%%%%%%%%%%%%%%%%%%%%%%%%%%%
\chapter{Introduction}\label{ch:intro}
%%%%%%%%%%%%%%%%%%%%%%%%%%%%%%%%%%%%%%%%%%%%%%%%%%%%%%

%%%%%%%%%%%%%%%%%%%%%%%%%%%%%%%%%%%%%%%%%%%%%%%%%%%%%%
\section{Motivation}
%%%%%%%%%%%%%%%%%%%%%%%%%%%%%%%%%%%%%%%%%%%%%%%%%%%%%%

Wormholes are hypothetical objects that represent shortcuts through spacetime, allowing observers to traverse between different regions \cite{Lobo2017,Visser:1995cc,Thorne1994}. The key feature that permits this behaviour is the presence of a minimal radius referred to as the "throat", that black holes do not present \cite{DInverno}, which removes the event horizon from the geometry of this object. As of now, these constructions remain theoretical, thus to study them one must rely on a solid theory that has withstood the test of time and has been rigorously verified through experimental verification. Currently, the best description that meets these requirements, and has been used to describe all known gravitational phenomena to date, is the General Relativity theory formulated by Einstein in 1915 \cite{Einstein1915}.

One may argue that the existence of wormholes is a subject of great speculation. While it is generally believed that these structures could potentially form in regions with extremely intense gravitational fields \cite{Visser:1995cc}, there is currently no concrete experimental evidence confirming or refuting their existence. Despite this lack of evidence, it is important to emphasize that it should not discourage the pursuit of this knowledge. In fact, even if wormholes may not naturally occur in the universe, they could be artificially created by an advanced civilization, much like how heavy elements of the periodic table are not found in nature but created artificially. The question of how one could create them is still far ahead; the initial step is to understand if these geometries are allowed in General Relativity.

In fact, this study not only aims to understand the possible existence of wormholes, but also delves into the foundations of General Relativity itself. On one hand, since General Relativity has undergone extensive testing and gained widespread acceptance \cite{Campbell1909}-\cite{Bertotti2003}, if it predicts the existence of physically plausible wormholes, then it justifies this direct study of these objects, as they may indeed exist or be possibly created. On the other hand, if General Relativity yields solutions that are physically implausible, then it provides valuable insights into the theory itself. It could lead to the conclusions that either wormhole geometries do not exist or General Relativity is not the best description of gravity. 

This last point may seem harsh at first glance, but let us remember that, despite its astonishing success, General Relativity is a scientific theory, and therefore, it is possibly provisional. This inherent condition of a scientific theory allows for scientific progress, as new information, whether experimental or not, may eventually arise and challenge the predictions stated by the theory, indicating the need for it to be reformulated or abandoned. In fact, this is already evident with GR, since phenomena such as the acceleration of the universe (leading to the introduction of dark energy) \cite{Lobo:2008sg}, the rotation curves of galaxies and the mass discrepancy of galaxy clusters (supporting the existence of dark matter) \cite{Lobo:2008sg}, and at the fundamental level, GR's resistance to quantization attempts \cite{Woodard:2009ns}, suggest that it should be reformulated. Overall, this motivates the study of modified theories of gravity (see Ref.\cite{Clifton:2011jh} for a comprehensive survey of modified theories and their cosmological consequences). In the particular case of wormholes, it is known that static traversable wormhole solutions violate the null energy condition (NEC) at the throat \cite{Morris:1988cz}, requiring the use of exotic matter to sustain these geometries. However, this matter has not been observed in astrophysical systems, which again prompts the consideration of modified gravity theories, where modifications in the gravitational sector may alleviate this need \cite{Lobo:2009ip}-\cite{Lobo:2007qi}.

Additionally, modifying a theory can be one of the most effective methods to gain a deeper understanding of it. To do so, two approaches can be employed. One can either question the foundational principles on which the theory is built, evaluate the importance of each one, and then construct a different consistent theory that omits one or a combination of these principles. This way, this approach aims to uncover a more fundamental theory constructed from the most basic principles. On another hand, one can adopt a broader perspective and explore whether the theory can be considered as part of a more general class of theories. In this second approach, which is the one employed in this thesis, General Relativity is treated as the starting point and one must choose theories that serve as tools to investigate how far, and in which direction, one can deviate from General Relativity \cite{Clifton:2011jh}. This way, these modified gravity theories must be simple and manageable, deviating from General Relativity in only one aspect, and their viability is discussed when applied to specific gravitational or cosmological phenomena. Overall, this is a trial-and-error approach to modifying gravity and should be seen as "gedanken-experiments" for a theoretician to probe the foundations of General Relativity.

Therefore, our goal is to develop new and physically plausible traversable wormhole solutions, specifically within the framework of $f(R,\mathcal{T})$ modified gravity theory (where $R$ is the Ricci scalar and $\mathcal{T}=T_{ab}T^{ab}$ with $T_{ab}$ being the energy-momentum tensor) \cite{katirci1,roshan1}, to provide new insights into the physics of these objects as well as into the modified gravity theory itself.

%%%%%%%%%%%%%%%%%%%%%%%%%%%%%%%%%%%%%%%%%%%%%%%%%%%%%%
\section{State-of-the-art}
%%%%%%%%%%%%%%%%%%%%%%%%%%%%%%%%%%%%%%%%%%%%%%%%%%%%%%

Now that we have provided a substantial motivation to delve into the wormhole physics, in this subsection, we begin by presenting the existing landscape in this field and outlining the contributions that our work brings to its advancement. Before doing so, let us establish a foundational understanding of what constitutes a wormhole. In essence, a wormhole is a topological entity that serves as a tunnel between two spacetime manifolds or distinct regions within the same spacetime manifold. In other words, wormholes manifest in at least two distinct varieties \cite{{Visser:1995cc}}:
\begin{itemize}
    \item Inter-universe wormholes: wormholes that connect "our" universe with "another" universe.
    \item Intra-universe wormholes: wormholes that connect two distant regions of "our" universe with each other.
\end{itemize}
In the scientific community, a linguistic ambiguity has surfaced among researchers that may confuse beginners. The term "multiverse" is commonly used to denote the conglomeration of many universes. In this context, a "universe" is defined as any sufficiently large and reasonable flat region of spacetime. Thus, inter-universe wormholes offer a way to traverse between different universes within the multiverse, while intra-universe wormholes provide a mechanism to traverse between two distant regions within a single universe.

Fortunately, the distinction between these two types of wormholes only becomes apparent when one examines the global topology. In the proximity of the wormhole's throat the physics remains indifferent to whether the travel is intra-universal or inter-universal, which implies that an observer making measurements near the wormhole's throat would be unable to differentiate between traveling to another universe or simply reaching a distant region within our own universe \cite{Visser:1995cc}. Computationally, this is advantageous, as it allows us to focus on the local dynamics near the wormhole's throat without an immediate concern for global considerations.

Let us now outline some of the work that has been done so far. The origins of wormhole physics can be traced back to Flamm's analysis of the Schwarzschild solution in 1916 \cite{Flamm}. Later, in 1935, Einstein and Rosen (ER) \cite{Einstein:1935tc} delved into wormhole-type solutions by constructing an elementary particle model, portraying a "bridge" that connects two identical sheets, which became known as an "Einstein-Rosen bridge" (ER bridge). In fact, ER aimed to construct a geometrical model that could represent a finite and singularity-free elementary "particle" through objects termed neutral and quasicharged bridges (for further generalizations of these constructions, see Ref.\cite{Visser:1995cc}). However, at the time, the distinction between "coordinate singularity" and "physical singularity" were not clear, with the event horizon commonly perceived as the singularity. 

Let us see this clearly through the example of the neutral ER bridge. This concept emerges from the observation that an appropriate coordinate transformation seems to eliminate the Schwarzschild (coordinate) singularity at $r=2M$, with $M$ being the mass of the object. This simply means that Einstein and Rosen identified a specific coordinate system that encompasses only two asymptotically flat regions of the maximally extended Schwarzschild spacetime. The standard Schwarzschild geometry \cite{Lobo:2017ctj}, given in spherical coordinates $x^a = \{t,r,\theta,\varphi\}$, is
\begin{equation}
    ds^2 = - \left(1-\frac{2M}{r} \right) dt^2 + \left(1-\frac{2M}{r} \right)^{-1} dr^2 + r^2 d \Omega^2,
\end{equation}
where $d\Omega^2 = d\theta^2 + \sin{\theta}^2 d\varphi^2$ is the solid angle surface element. Performing a coordinate change $u^2=r-2M$, the line element is represented by the ER form 
\begin{equation}
    ds^2 = - \frac{u^2}{u^2 + 2M}dt^2 + 4 \left(u^2 + 2M\right)du^2 + \left(u^2 + 2M\right) d \Omega^2,
\end{equation} 
where $u \in (-\infty,+\infty)$. It is important to highlight that this coordinate transformation deliberately excludes the region containing the curvature singularity, corresponding to $r \in [0,2M)$. The vicinity around $u=0$ is conceptualized as a "bridge", connecting the asymptotically flat regions near $u=+\infty$ and $u=-\infty$.

To justify the "bridge" designation, let us examine a spherical surface with a constant $u$ value, where the area is given by $A(u) = 4\pi(2M + u^2)^2$. This area reaches its minimum at $u=0$, denoted as the "throat" in contemporary language, with $A(0) = 4\pi(2M)^2$. The surrounding area is commonly termed the "bridge" or the "wormhole". As a result, the neutral ER bridge, or the Schwarzschild wormhole, constitutes a portion of the maximally extended Schwarzschild geometry. However, this wormhole is nontraversable, as the throat pinches off before any observer can traverse it \cite{Visser:1995cc}. 

As for the quasi-charged ER bridge, one starts with the Reissner-Nordstr\"om metric, which describes an electrically charged black hole characterized by charge $Q$ and mass $M$. After a particular coordinate transformation \cite{Lobo:2017ctj}, one obtains a geometry that represents a massless, quasicharged entity characterized by negative energy density, featuring a horizon at $u=0$. This is the very object that ER aimed to interpret as an "electron". However, as before, it is merely a coordinate artifact arising from the adoption of a specific coordinate patch intentionally crafted to double-cover the asymptotically flat region beyond the black hole event horizon, and therefore, provides no protection from potential threats if one were to make the ill-advised attempt to traverse it.

In the 1950s, after a period of inactivity, Wheeler revived interest in wormholes by conceptualizing them as quantum foam entities operating at the Planck scale \cite{Wheeler:1955zz, Wheeler1962}. However, these structures, known as Wheeler wormholes or "geons", were non-traversable and, in principle, would develop some form of singularity \cite{Geroch:1967fs}. Despite the exploration by many physicists in different contexts \cite{Ernst:1957zza}-\cite{Diemer:1999hy}, the ambitious nature and the lack of experimental evidence posed challenges, leading to a decline in interest.

Later, classical wormholes had been theorized and explored \cite{Ellis:1973yv}-\cite{Clement}, but it wasn't until the groundbreaking paper by Morris and Thorne in 1988 \cite{Morris:1988cz} that the field gained broad attention. In this paper, the authors recognized the lack of feasible technology for traversable wormholes, opting instead to address the fundamental question: "What do the laws of physics permit?". This led to a different approach in solving the Einstein field equations. Rather than starting with a known matter source and computing the resulting spacetime metric, they reversed this strategy. They began by considering a specific spacetime metric and then computed the corresponding matter source that would give rise to such geometry.

In this manner, they discovered a unique property in some of these solutions known as "exotic matter". This involves a energy-momentum tensor that violates the null energy condition (NEC), which stipulates that for any null vector $k^a$, the energy-momentum tensor $T_{ab}$ must obey to $T_{ab}k^ak^b \geq 0$. In fact, the null energy condition is the weakest of the energy conditions. Once violated, the other energy conditions are also violated: the weak energy condition (WEC) that asserts that for any timelike vector $U^a$, $T_{ab}U^aU^b \geq 0$; the strong energy condition (SEC) that states that for any timelike vector $U^a$, $\left(T_{ab} -\frac{T}{2}g_{ab}\right)U^a U^b \geq 0$, where $T$ is the trace of the energy-momentum tensor and $g_{ab}$ is the metric; the dominant energy condition (DEC) that requires that for any timelike vector $U^a$, $T_{ab}U^aU^b \geq 0$ and $T_{ab}U^b$ is not spacelike. In general, classical forms of matter are believed to adhere to these energy conditions \cite{HawEllis}, while certain quantum fields, exemplified by phenomena such as the Casimir effect and Hawking evaporation \cite{Klinkhammer:1991ki}, have been found to violate them.

It was further found that in quantum systems within classical gravitational backgrounds, violations of the weak or null energy conditions are typically small and compensated by subsequent positive energy states. To accommodate pointwise energy condition violations, researchers have introduced averaged energy conditions over timelike or null geodesics \cite{Tipler:1978zz,Roman:1986tp}. This framework permits localized violations as long as the conditions are satisfied when averaged along a null or timelike geodesic \cite{Tipler:1978zz}. 

Pioneering work by Ford in the late 1970's introduced a new set of energy constraints \cite{Ford:1978qya}, which later evolved into constraints on negative energy fluxes in 1991 \cite{Ford:1990id}. These constraints eventually took the form of the Quantum Inequality (QI) applied to energy densities, as introduced by Ford and Roman in 1995 \cite{Ford:1994bj}-\cite{Pfenning:1996tb}. When applied to curved spacetimes, the QI implies that these spacetimes appear flat within sufficiently small regions. In the context of wormhole geometries \cite{Ford:1995wg,Kuhfittig:2002ur}, considering a small spacetime volume around the throat of the wormhole, with dimensions much smaller than the minimum proper radius of curvature, allows for treating the spacetime as approximately flat in this region and applying the QI constraint, led to the conclusion that traversable wormholes would either have extremely small throats (slightly larger than the Planck length) or exhibit significant discrepancies in the length scales characterizing the wormhole geometry. This analysis implies that exotic matter is generally confined to an extremely thin band, and /or significant red-shifts are involved, from which Ford and Roman concluded that the existence of macroscopic traversable is highly improbable \cite{Ford:1995wg,LoboCrawford,Roman:2004xm}. 

However, an alternative perspective arises when considering a specific form of exotic matter that does not originate from quantum field theory. Recent cosmological observations have confirmed the Universe's accelerated expansion \cite{SupernovaSearchTeam:2001qse}-\cite{WMAP:2003zzr}, attributed to dark energy comprising approximately $70\%$ of the cosmic composition \cite{WMAP:2003ivt}. Various candidates for dark energy, including a positive cosmological constant, quintessence fields, generalized Chaplygin gas, and tachyon models, have been proposed. Characterized by the equation of state $\omega = p/\rho$, where $p$ is pressure and $\rho$ is energy density, values of $\omega < -1/3$ drive cosmic expansion, with $\omega = -1$ corresponding to a cosmological constant. The one that is particularly interesting for us is the concept of "phantom energy" \cite{Melchiorri:2002ux,Alcaniz:2003qy,Carroll:2003st}, denoted by $\omega < -1$, since it violates the NEC, thus providing a basis for exploring wormhole physics. Wormholes supported by phantom energy \cite{Lobo:2005us}-\cite{Kuhfittig:2006xj}, generalized Chaplygin gas \cite{Lobo:2005vc}, and the van der Waals equation of state \cite{Lobo:2006ue} have been subjects of investigation. Additionally, this accelerated cosmic expansion permits the growth of macroscopic wormholes from submicroscopic quantum foam structures (see Ref.\cite{Gonzalez-Diaz:2003xgh}-\cite{Gonzalez-Diaz:2003xmx} for more research done in this field). 

Overall, the exotic matter poses significant challenges in wormhole physics. Thus, the primary focus on this field has been on minimizing violations of the null energy condition, a common issue encountered in static wormholes \cite{Visser:1995cc,Morris:1988cz}. Morris and Thorne initially addressed this challenge by constructing specific wormhole geometries to reduce the region of violation \cite{Morris:1988cz}. Later, Visser \cite{Visser:1989kh} proposed different geometries, where the exotic matter is concentrated only at the edges and corners. This design allows a traveler to pass through the flat faces without encountering any matter. Kuhfittig further demonstrated that the region of exotic matter can be made arbitrarily small \cite{Kuhfittig:1999nd,Kuhfittig:2003pu}. In dynamic wormholes, specifically concerning the averaged null energy condition, certain regions can be identified to avoid violations \cite{Visser:1997yn}-\cite{Kim:1995xf}. More recently, Visser and colleagues introduced a way to quantify the "total amount"of energy condition-violating matter, using a "volume integral quantifier" \cite{Visser:2003yf,Kar:2004hc}. This involves calculating definite integrals, such as $\int T_{ab}U^a U^b dV$ and $\int T_{ab} k^a k^b dV$, where the extent of the violation is determined by the negativity of these integrals. While null energy and averaged null energy conditions are always violated in wormhole spacetimes, specific examples explored by Visser et al. demonstrated that wormholes can be supported by arbitrarily small quantities of averaged null energy condition-violating matter.

Nevertheless, it is important to recognize that the landscape of energy conditions has witnessed a substantial transformation, particularly with the revelation that classical systems can violate all energy conditions \cite{Barcelo:2000zf}. Moreover, recent cosmological observations suggest potential violations of the SEC as well as hint at possible violations of NEC in classical regimes \cite{SupernovaSearchTeam:2004lze,Visser:2003vq,Caldwell:2003vq}, therefore casting doubt on the once-unquestioned status of energy conditions as fundamental laws \cite{Barcelo:2002bv}.

When it comes to experimental evidence, the existence of wormholes remains elusive. Despite the absence of direct observations, theorists are exploring potential avenues to bridge theoretical concepts with observable data. Given the prevalence of stars and black holes in the universe, there is a hypothesis that wormholes, whether naturally occurring or artificially constructed, might exist in substantial numbers. If these wormholes inhabit cosmological space, they could manifest microlensing effects on point sources at both non-cosmological \cite{Cramer:1994qj} and cosmological distances \cite{Torres:1998cu,Safonova:2001vz}, potentially influencing phenomena like gamma-ray bursts. Wormholes of significant size might even induce macrolensing effects \cite{Safonova:2001nd}. The emerging consensus is that there is a conceptual link between wormholes, stars, and black holes. Works carried in Refs. \cite{Gonzalez-Diaz:1996iea,DeBenedictis:2000rg,Visser:1997yn,Hayward:2002sz,Hochberg:1998ha} suggest that there is a continuum of objects ranging from stars to wormholes through black holes, where stars consist of normal matter, black holes of vacuum, and wormholes of exotic matter. In a broader sense, wormholes could be referred as "exotic stars".

In pursuit of our objectives we adopt a different approach that has not yet been discussed in this section. Specifically, to overcome the use of exotic matter, we consider wormhole solutions within the framework of modified theories of gravity \cite{Agnese:1995kd}-\cite{lobo4}, where we introduce additional components in the gravitational sector to maintain the traversability of the wormhole throat's geometry while ensuring that the matter components remain non-exotic. This outcome can be achieved through various modifications of GR, ranging from $f(R)$ gravity and its extensions \cite{Lobo:2009ip}-\cite{Rosa:2022osy} to scenarios involving couplings between curvature and matter \cite{Garcia:2010xb,MontelongoGarcia:2010xd}, theories featuring additional fundamental fields \cite{Harko:2013yb,Anchordoqui:1996jh}, Gauss-Bonnet gravity \cite{Bhawal:1992sz,Dotti:2007az,Mehdizadeh:2015jra}, and braneworld scenarios \cite{Bronnikov:2002rn,Lobo:2007qi}.

To obtain physically relevant spacetime solutions describing localized objects, the so-called junction condition are usually employed using the distribution formalism. These junction conditions are theory-dependent, therefore must be derived independently for each theory. Regarding General Relativity they were derived long ago \cite{Israel:1966rt} and have been applied in various astrophysical systems, such as the analysis of traversable wormholes \cite{Lobo:2005yv}-\cite{Visser:1989kh}, fluid stars \cite{schwarzschild1,rosafluid,Rosa:2023hfm}, and gravitational collapse \cite{oppenheimer1,rosa112}. For modified theories of gravity, junction conditions have been derived for $f(R)$ theories and its extensions \cite{senovilla1}-\cite{rosafrt}, theories with additional fundamental fields \cite{rosafrt2,suffern,Barrabes:1997kk} and metric-affine gravity \cite{Padilla:2012ze}-\cite{delaCruz-Dombriz:2014zaa}.

In particular, a well-explored extension to General Relativity is the $f(R)$ gravity theory, where the Ricci scalar $R$ in the Einstein-Hilbert action of General Relativity is replaced by a function of $R$. This theory has demonstrated its effectiveness in accounting some cosmological phenomena such as the accelerated expansion of the Universe \cite{Elizalde:2010ts} and offered to be a potential alternative to dark matter \cite{Capozziello:2006uv,Boehmer:2007kx}. However, certain $f(R)$ models have been ruled out based on weak-field limit tests \cite{Capozziello:2007ms,Olmo:2006eh} and the presence of matter instabilities \cite{Nojiri:2010wj}, suggesting that it may not be the most accurate description of gravity. In the context of wormholes, Ref. \cite{Mishra:2021xfl} revealed that a wormhole model based on $f(R)$ theory satisfies only the transverse null energy condition, indicating that the challenge of eliminating exotic matter still persists.

Due to these challenges, additional extensions have been explored. One notable example is the $f(R,T)$ theory, where the function of $R$ in the theory's action is substituted with a function of both $R$ and the trace of the energy-momentum tensor $T$. This theory has demonstrated utility in various astrophysical scenarios \cite{Harko:2011kv,Zaregonbadi:2016xna,Dey:2020skl}, although it has encountered limitations in consistently aligning with observational data in some cases \cite{Velten:2017hhf}. Within this framework, researchers have examined wormhole models \cite{Banerjee:2020uyi,Dixit:2020nmn,Rosa:2022osy}, and specifically in Ref. \cite{Banerjee:2020uyi} solutions were identified that satisfy the NEC throughout the spacetime without the need for thin-shell formalism, while in Ref. \cite{Rosa:2022osy} the authors have obtained solutions without the thin-shell formalism as well but satisfying all the energy conditions (NEC, WEC, SEC and DEC) throughout the spacetime this time.

In this thesis, our focus lies on a covariant generalization of Einstein’s General Relativity known as energy-momentum squared gravity, or $f\left(R, \mathcal{T} \right)$ gravity \cite{Katirci:2013okf,Roshan:2016mbt}, where $\mathcal{T}=T_{ab}T^{ab}$. This modification shares similarities with previously explored curvature-matter coupling theories \cite{Harko:2011kv,Bertolami:2007gv,Harko:2010mv,Haghani:2013oma}, where the energy-momentum tensor is not conserved. While $f\left(R, \mathcal{T} \right)$ gravity has been extensively studied in various contexts, ranging from cosmological models \cite{Chamel:2016ynd}-\cite{Barbar:2019rfn} to compact objects \cite{Nari:2018aqs}-\cite{Sharif:2021uyc}, including black holes \cite{Chen:2019dip,Rudra:2020rhs}, the literature on wormhole physics within this framework is limited. Some specific solutions have been identified through a Noether symmetry approach; however, these solutions lack physical relevance as they violate the NEC and, consequently, all the other more restrictive energy conditions \cite{Sharif:2021gdv,ZeeshanGul:2023ysx}. Therefore, the primary objective of this work is thus to address this gap in the literature by providing a comprehensive analysis of physically relevant traversable wormhole spacetimes in $f\left(R, \mathcal{T} \right)$ gravity. Consequently, to obtain physically reasonable solutions, our second goal of this work is then to provide the junction conditions of linear $f\left(R, \mathcal{T} \right)$ as well as explicit examples of their application.

This thesis is structured as follows: In Chapter \ref{ch:framework}, we provide an introduction to the General Relativity theory, along with a derivation of the junction conditions within GR. This foundation sets the stage for the analysis of wormholes in GR, encompassing their metric, field equations, and an examination of the matter components with respect to the energy conditions. This way, from Chapter \ref{ch:framework}, we understand the work done by Morris and Thorne \cite{Morris:1988cz}, identifying the challenges, particularly the reliance on exotic matter to sustain these geometric structures. Moving forward, Chapter \ref{ch: MGT} introduces modified gravity theories, focusing on the $f(R,\mathcal{T})$. We derive the field equations for wormholes within the framework of a linear $f(R,\mathcal{T})$ function, and in order to obtain physically relevant solutions, we derive the junction conditions and the matching process to a vacuum exterior. Furthermore, we extend this investigation to higher powers of $\mathcal{T}$, deriving the corresponding field equations and junction conditions to gain additional insights into these solutions. In the following chapter \ref{ch:results}, we showcase specific examples of the derived wormhole solutions. This encompasses both the linear scenario, characterized by $f(R,\mathcal{T})=R+\gamma \mathcal{T}$, and the quadratic scenario, defined by $f(R,\mathcal{T})=R+\sigma \mathcal{T}^2$. Additionally, we present the outcomes resulting from the matching process with a vacuum exterior for both cases. Lastly, Chapter \ref{ch:conclusion} encapsulates the concluding remarks and insights derived from this study.

%%%%%%%%%%%%%%%%%%%%%%%%%%%%%%%%%%%%%%%%%%%%%%%%%%%%%%
\chapter{Theoretical framework}\label{ch:framework}
%%%%%%%%%%%%%%%%%%%%%%%%%%%%%%%%%%%%%%%%%%%%%%%%%%%%%%

%%%%%%%%%%%%%%%%%%%%%%%%%%%%%%%%%%%%%%%%%%%%%%%%%%%%%%
\section{Introduction to General Relativity}
%%%%%%%%%%%%%%%%%%%%%%%%%%%%%%%%%%%%%%%%%%%%%%%%%%%%%%

In this chapter, we start by reviewing General Relativity. Its astonishing success and extensive experimental validation have established it as the core of our understanding of gravity, so much so, that modified theories of gravity often find their roots as adjustments or extensions of Einstein's General Relativity theory. After deriving the Einstein field equation, we derive the junction conditions using the distribution formalism. Consequently, through the Einstein field equation and wormhole metric, we obtain the matter quantities and analyse their meaning in this context. 

\subsection{The role of geometry}

The key ingredient of General Relativity is to treat the geometry as the description of the spacetime through differential geometry in particular. The basic concept from which differential geometry is built is the manifold, which is a set that locally looks like $\mathbb{R}^n$, i.e. resembles Euclidean spacetime (for a rigorous definition of the manifold and discussion on differential geometry see \cite{Wald1984,DInverno,HawEllis,Misner1973}). Thus, spacetime is treated as a 4-dimensional manifold, where one defines a symmetric non-degenerate metric $g_{ab}$, i.e. the determinant of the metric $g$ is non-zero, and a quantity related to parallel transport (transport of a vector along a curve while maintaining its alignment with respect to the underlying space) $\Gamma^c_{ab}$ known as the connection (for an introduction see Ref.\cite{DInverno} and \cite{HawEllis} for a more advanced treatment). Furthermore, this connection leads to the definition of a derivative adapted to the curved manifold, known as covariant derivative, denoted by $\nabla$, which definition is
\begin{equation}
    \nabla_c T^a_b = \partial_c T^a_b + \Gamma^a_{dc} T^d_b - \Gamma^d_{bc} T^a_d ,
\end{equation}
for a given tensor $T^a_b$. In fact, in General Relativity this connection $\Gamma^c_{ab}$ is associated with the metric $g_{ab}$ and is referred as the Levi-Civita connection (see Ref. \cite{Wald1984} for a derivation of the Levi-Civita connection and \cite{Shrodinger} for more general connections). 

Additionally, in the particular case of GR, where $T_{ab}$ is the energy-momentum tensor and $\nabla_c$ is a covariant derivative derived from the Levi-Civita connection associated with the metric, this implies that $\nabla_c T^{ab}=0$. To fulfill this requirement, it turns out that one needs two things. Firstly, the connection must be symmetric regarding its two lower indices
\begin{equation}
    \Gamma^a_{bc}=\Gamma^a_{cb}.
    \label{torsion-less}
\end{equation}
And secondly, the metric must be covariantly conserved
\begin{equation}
    \nabla_c g_{ab}=0.
    \label{nullmetricity}
\end{equation}
With these choices, the connection then takes the Levi-Civita form 
\begin{equation}
    \Gamma^a_{bc}=\left\{^{a}_{bc}\right\}=\frac{1}{2}g^{ad}\left[\partial_b g_{cd} + \partial_c g_{db} - \partial_d g_{bc}\right] ,
    \label{LeviCivita}
\end{equation}
where one can see explicitly its relation to the metric $g_{ab}$.

Furthermore, the curvature of the manifold is described by the Riemann tensor, which can be derived from the connection in the following manner
\begin{equation}
    R^a_{bcd}=\partial_c \Gamma^a_{bd}-\partial_d \Gamma^a_{bc}+\Gamma^e_{bd} \Gamma^a_{ec} - \Gamma^e_{bc} \Gamma^a_{ed} .
\end{equation}
Additionally, contracting the first and third indices gives us the Ricci tensor
\begin{equation}
    R_{ab}=R^c_{acb}=-R^c_{abc}.
\end{equation}
And, using the metric to contract with the Ricci tensor, we obtain the Ricci scalar
\begin{equation}
    R = g^{ab}R_{ab}.
\end{equation}

\subsection{General Relativity}

Now, to describe the dynamics of the gravitational field one can remain on a close analogy to the Poisson equation, $\nabla^2 \phi=4\pi \rho$, which describes the dynamics of the gravitational potential in Newtonian gravity. In fact, General Relativity equations in empty space are simply $R_{ab}=0$, where this Ricci tensor is constructed from the Levi-Civita connection \eqref{LeviCivita}. Therefore, noticing that the Ricci tensor is a second order differential expression with respect to the metric, one can see that $R_{ab}=0$ stands as a good analogy with Laplace's equation $\nabla^2 \phi = 0$.

Moreover, to extend this analogy to the case where we have matter fields we must introduce some extra assumptions. Firstly, the only field of the General Relativity theory is the metric one, which means that all the additional fields are considered to be "matter fields", that act as sources for the "gravitational field". Therefore, gravity is associated only to the second rank tensor field that represents the metric, and the field equation must present a left side depending only on the metric and the right side containing the dependence of all the other fields. The tensor that characterizes and plays the role of the source of the gravitational field is the energy-momentum tensor $T_{ab}$ (Ref. \cite{Misner1973} presents a detailed discussion, while \cite{ Buchdahl,Sotiriou:2007zu} shows some problems related to the definition of this quantity). Additionally, since we are looking for an analogy with the Poisson's equation, then our field equation must be a second order differential equation.

In summary, these are the necessary assumptions of General Relativity: 
\begin{itemize}
    \item The connection must be symmetric in its two lower indices $\Gamma^a_{bc}=\Gamma^a_{cb}$;
    \item The metric is conserved $\nabla_c g_{ab}=0$;
    \item The gravitational field is associated with the metric and no other fields are involved in this interaction;
    \item The field equations should be second order partial differential equations; %isto tem um nome fancy penso eu
    \item The field equations should be covariant.
\end{itemize}

Now, one can follow the original Einstein's derivation to obtain the field equations \cite{Wald1984,Misner1973,Bernard1985}, given by
\begin{equation}
    G_{ab}= R_{ab}-\frac{1}{2}R g_{ab}= \kappa T_{ab},
\end{equation}
where $\kappa=8 \pi G/c^4$, with $G$ being the gravitational constant and $c$ the speed of light. To simplify the notation, in this thesis we adopt a system of geometrized units for which $G=c=1$, thus $\kappa= 8 \pi$.

\subsection{The Lagrangian formulation of General Relativity}

The Einstein field equations can be derived through a Lagrangian formulation, which proves to be useful for modifying General Relativity and comparing different modified theories. Let us then develop this method in this section.  

To apply the Lagrangian formulation, we must remember that, as established in the previous section, the metric $g_{ab}$ is the fundamental field describing the gravitational dynamics. Therefore, it plays the role of a dynamical variable. To construct the gravitational action, we must first establish some assumptions. The first one states that the action must be a generally covariant scalar. Since the volume element is a tensor density of weight $-1$ (see \cite{DInverno} for a definition), this leads us to include a multiplicative factor of $\sqrt{-g}$ in the action, where $g$ is the determinant of the metric tensor $g_{ab}$. In order to obtain second-order partial differential equations, the second assumption states that the action must depend only on the metric and its first derivatives. 

In practice, the Ricci scalar $R$ is the simplest scalar quantity constructed with the metric, that involves second derivatives of $g_{ab}$. In fact, there is no scalar quantity constructed solely with the metric and its first derivative. This limitation arises from the fact that these are not covariant objects, and thus, no combination of them would yield a covariant result. Taking these assumptions into account, we define the gravitational action, known as the Einstein-Hilbert action, as follows
\begin{equation}
    S_{EH}=\frac{1}{16 \pi} \int_U \sqrt{-g} R d^4x . 
\end{equation}
The application of the variational principle leads to
\begin{equation}
    \delta S_{EH}=\delta \int_U \sqrt{-g} R d^4x =0,
\end{equation}
which can be grouped in three terms
\begin{equation}
    \underbrace{\int_U \delta \left(\sqrt{-g}\right) g_{ab} R^{ab} d^4x}_{(I)} +\underbrace{\int_U \sqrt{-g} \delta \left(g_{ab}\right) R^{ab} d^4 x}_{(II)}+\underbrace{\int_U \sqrt{-g} g_{ab} \delta R^{ab} d^4 x}_{(III)}.
    \label{3terms}
\end{equation}
Let us start by computing the third term related to the variation of the Ricci tensor. To do so, we begin by writing the Ricci tensor explicitly in terms of the Levi-Civita connection $\Gamma^c_{ab}$ given by
\begin{equation}
    R_{ab}=R^c_{acb}=\partial_c \Gamma^c_{ab}-\partial_b \Gamma^c_{ac}+\Gamma^e_{ab}\Gamma^c_{ec}-\Gamma^e_{ac}\Gamma^c_{eb}.
\end{equation}
Taking now the variation of $R_{ab}$ results in
\begin{equation}
    \delta R_{a b}=\partial_c \delta \Gamma_{a b}^c-\partial_b \delta \Gamma_{a c}^c+\delta \Gamma_{a b}^e \Gamma_{e c}^c+\Gamma_{a b}^e \delta \Gamma_{e c}^c-\delta \Gamma_{a c}^e \Gamma_{e b}^c-\Gamma_{a c}^e \delta \Gamma_{e b}^c .
    \label{deltaRab}
\end{equation}
To simplify this result let us take the covariant derivative of the Levi-Civita connection $\Gamma^a_{bc}$ given as follows
\begin{equation}
    \nabla_d \Gamma^a_{bc}=\partial_d \Gamma^a_{bc} + \Gamma^a_{de} \Gamma^e_{bc} - \Gamma^e_{bd}\Gamma^a_{ec} - \Gamma^e_{cd} \Gamma^a_{be},
\end{equation}
where we can substitute the $\Gamma^a_{bc}$ tensor by $\delta \Gamma^a_{bc}$ and obtain
\begin{equation}
    \nabla_d \delta \Gamma^a_{bc} = \partial_d \delta \Gamma^a_{bc} + \Gamma^a_{de}\delta \Gamma^e_{bc} - \Gamma^e_{bd}\delta \Gamma^a_{ec} - \Gamma^e_{cd} \delta \Gamma^a_{be}
\end{equation}
This way we can identify the terms in Eq.\eqref{deltaRab} with the two following covariant derivatives of the Levi-Civita connection $\Gamma^c_{ab}$
\begin{equation}
    \nabla_c \delta \Gamma^c_{ab} = \partial_c \delta \Gamma^c_{ab} + \Gamma^c_{ce} \delta \Gamma^e_{ab} - \Gamma^e_{ac} \delta \Gamma^c_{eb} - \Gamma^e_{bc} \delta \Gamma^c_{ae},
    \label{ccab}
\end{equation}
\begin{equation}
    \nabla_b \delta \Gamma^c_{ac} = \partial_b \delta \Gamma^c_{ac} + \Gamma^c_{be} \delta \Gamma^e_{ac} - \Gamma^e_{ab} \delta \Gamma^c_{ec} - \Gamma^e_{cb} \delta \Gamma^c_{ae}.
    \label{bcac}
\end{equation}
To see it explicitly, let us rewrite Eqs.\eqref{ccab} and \eqref{bcac} as
\begin{equation}
    \partial_c \delta \Gamma^c_{ab} + \Gamma^c_{ce} \delta \Gamma^e_{ab} = \nabla_c \delta \Gamma^c_{ab} + \Gamma^e_{ac} \delta \Gamma^c_{eb} + \Gamma^e_{bc} \delta \Gamma^c_{ae},
\end{equation}
\begin{equation}
    -\partial_b \delta \Gamma^c_{ac} - \Gamma^c_{be} \delta \Gamma^e_{ac} + \Gamma^e_{ab} \delta \Gamma^c_{ec} = - \nabla_b \delta \Gamma^c_{ac} - \Gamma^e_{cb} \delta \Gamma^c_{ae}.
\end{equation}
Now it is clear that the terms on the left hand side of these equations are the ones that appear in Eq.\eqref{deltaRab}. Therefore, substituting these results into Eq.\eqref{deltaRab} we get
\begin{equation}
    \delta R_{ab} = \nabla_c \delta \Gamma^c_{ab} + \Gamma^e_{ac} \delta \Gamma^c_{eb} + \Gamma^e_{bc} \delta \Gamma^c_{ae} - \nabla_b \delta \Gamma^c_{ac} - \Gamma^e_{cb} \delta \Gamma^c_{ae} - \Gamma^e_{ac} \delta \Gamma^c_{eb}.
\end{equation}
Remembering the fact that the Levi-Civita connection is symmetric, $\Gamma^c_{ab}=\Gamma^c_{ba}$, this result simplifies to
\begin{equation}
    \delta R_{ab} = \nabla_c \delta \Gamma^c_{ab} - \nabla_b \delta \Gamma^c_{ac}.
\end{equation}
Therefore, the third term of Eq.\eqref{3terms} becomes
\begin{equation}
    \int_U \sqrt{-g} g_{ab} \delta R^{ab} d^4 x = \int_U \sqrt{-g} g^{ab} \delta R_{ab} d^4 x = \int_U \sqrt{-g} g^{ab} \left(\nabla_c \delta \Gamma^c_{ab} - \nabla_b \delta \Gamma^c_{ac} \right) d^4 ,
\end{equation}
which can be rewritten as
\begin{equation}
    \int_U \sqrt{-g} g_{ab} \delta R^{ab} d^4 x = \int_u \sqrt{-g} \nabla_c \left(g^{ab} \delta \Gamma^c_{ab} - g^{ac} \delta \Gamma^b_{ab}\right),
\end{equation}
where we applied the Leibniz rule to $\nabla_c \left(g^{ab} \delta \Gamma^c_{ab}\right)$, with the fact that the covariant derivative of the metric $g_{ab}$ vanishes by definition, i.e. $\nabla_c g_{ab}=0$, and manipulated the indices of the right term.
Defining a new vector $A^c \equiv g^{ab} \delta \Gamma^c_{ab} - g^{ac} \delta \Gamma^b_{ab}$, this integral becomes
\begin{equation}
    \int_U \sqrt{-g} g_{ab} \delta R^{ab} d^4 x = \int_U \sqrt{-g} \nabla_c A^c d^4x.
\end{equation}
Now, using the fact that the covariant derivative of a vector field $A^a$ is given by \cite{Rosa:2019ykz}
\begin{equation}
    \nabla_a A^a = \frac{1}{\sqrt{-g}} \partial_a \left(\sqrt{-g} g^{ab} A_b\right),
\end{equation}
the integral takes the form
\begin{equation}
    \int_U \sqrt{-g} g_{ab} \delta R^{ab} d^4 x = \int_U \sqrt{-g} \frac{1}{\sqrt{-g}} \partial_c \left(\sqrt{-g} g^{cb} A_b\right) d^4 x = \int_U \partial_c \left(\sqrt{-g} A^c\right) d^4 x.
\end{equation}
Applying the stokes theorem to this result leads us to
\begin{equation}
    \int_U \sqrt{-g} g_{ab} \delta R^{ab} d^4 x = \int_{\partial U} \sqrt{-g} A^c d^3 x = 0,
\end{equation}
since the integral must vanish at the boundary $\partial U$, which in this case refers to infinity. Therefore, we have shown that the third term of Eq. \ref{3terms} is null, and thus does not contribute to the equation of motion for the metric tensor. 

Now, to compute the first two terms we use the fact that the inverse of the metric tensor is related to the metric by $\delta^a_b=g^{ac}g_{cb}$, where $\delta^a_b$ is the Kronecker delta tensor whose entries are equal to $1$ if $a=b$ and $0$ if $a\neq b$. Taking the variation of this expression gets us 
\begin{equation}
    0 = \delta \left(g^{ac}g_{cb} \right) = \delta \left(g^{ac}\right) g_{cb} + g^{ac}\delta \left(g_{cb}\right),
\end{equation}
which results into
\begin{equation}
    \delta g^{ab} = - g^{ac}g^{bd} \delta g_{cd}.
    \label{partial gab}
\end{equation}
Furthermore, to compute the first term, we introduce the definition of the inverse of a general matrix $[a_{ij}]$, which is given as 
\begin{equation}
    \left[a_{ij}\right]^{-1}=\frac{1}{a} \text{adj}\left[a_{ij}\right],
    \label{inversematrix}
\end{equation}
where, $a$ and $\text{adj}\left(a_{ij}\right)$, are the determinant and the adjoint matrix, respectively. The determinant is obtained using the Laplace's rule as follows
\begin{equation}
    a = \sum^n_{j=1} a_{ij} \text{adj}\left[a_{ij}\right],
\end{equation}
from which we can derive the following partial derivative
\begin{equation}
    \frac{\partial a}{\partial a_{ij}} = \text{adj} \left[a_{ij}\right].
    \label{da/daij}
\end{equation}
Considering now that $a$ is given by the composite function $a(a_{ij}(x^k))$, then its derivative is given by 
\begin{equation}
    \frac{\partial a}{\partial x^k}= \frac{\partial a}{\partial a_{ij}} \frac{\partial a_{ij}}{\partial x^k},
\end{equation}
which can be further simplified when Eq.\eqref{inversematrix} and \eqref{da/daij} is taken into account, leading to
\begin{equation}
    \frac{\partial a}{\partial x^k} = a \left[a_{ij}\right]^{-1} \frac{\partial a_{ij}}{\partial x^k}.
\end{equation}
Applying this result to the metric $g_{ab}$, using the fact that $\delta^a_b=g^{ac}g_{cb}$, gives us 
\begin{equation}
    \partial_a g = g g^{bc} \partial_a g_{bc}.
\end{equation}
Additionally, we can substitute $\partial$ by $\delta$, and $g$ by $\sqrt{-g}$ to obtain the following expression
\begin{equation}
    \delta \sqrt{\left(-g\right)} = \frac{1}{2} \sqrt{\left(-g\right)} g^{ab} \delta g_{ab} . 
    \label{delta sqrt g}
\end{equation}
Finally, we have everything to compute the term $(I) + (II)$ in equation \eqref{3terms}, through the results \eqref{partial gab} and \eqref{delta sqrt g}. Performing some algebraic manipulation the term $(I) + (II)$, therefore the Eq.\eqref{3terms} since the term $(III)$ is null,  is given by
\begin{equation}
    - \int_U \sqrt{\left(-g\right)} \left[R^{ab}-\frac{1}{2}g^{ab}R\right] \delta g_{ab}d^4x = 0.
\end{equation}
Using the fact that  $\delta g_{ab} \neq 0$, since this is considered as the independent variation, implies that the previous integral is null, if and only if the following is obeyed 
\begin{equation}
    G^{ab} \equiv R^{ab} - \frac{1}{2} g^{ab}R=0,
\end{equation}
where the Einstein tensor $G_{ab}$ was defined. Note that these are the Einstein field equations in the absence of matter. 

To take into account the source of the field, we must add another term to the action, the so-called matter term that is associated with the matter field. This matter action is defined as
\begin{equation}
    S_M = \int_U \sqrt{-g} \mathcal{L}_M \left(g_{ab},\psi\right) d^4x , 
\end{equation}
where $\mathcal{L}_M$ is the matter Lagrangian and $\psi$ collectively denotes the matter fields. Varying this matter action with respect to the metric defines the energy-momentum tensor
\begin{equation}
    T_{ab} = \frac{-2}{\sqrt{-g}} \frac{\delta\left(\sqrt{-g}\mathcal{L}_M\right)}{ \delta g^{ab}}.
\end{equation}
Then, when we finally consider the variation of the action $S=S_{EH}+S_M$, with respect to the metric, we obtain the full Einstein field equations
\begin{equation}
    G_{ab} = \kappa T_{ab}.
    \label{Einstein eq}
\end{equation}
Note that these are the equations that describe the dynamics of General Relativity. Therefore, any modification of this theory leads to a modification of the action, and thus, these equations.

%%%%%%%%%%%%%%%%%%%%%%%%%%%%%%%%%%%%%%%%%%%%%%%%%%%%%%
\subsection{Distributional formalism and junction conditions}\label{subsec:JC}
%%%%%%%%%%%%%%%%%%%%%%%%%%%%%%%%%%%%%%%%%%%%%%%%%%%%%%

Now, to obtain physically plausible wormhole solutions, one may require the use of the so-called junction conditions, which are derived using the distributional formalism. This approach allows one to match two existing spacetimes defined in complementary regions, such that the full solution is physically reasonable. Given the fact that these are theory-dependent, we initially derive them for the case of GR, and later, in Chapter \ref{ch:modifiedgravity}, we extend this derivation to the case of $f(R, \mathcal{T})$ modified gravity theory. 

This matching takes place on a 3-dimensional hypersurface $\Sigma$ with constant radius $r_\Sigma$, separating the interior spacetime in the $V^{-}$ region, which metric is denoted by $g^{-}_{ab}$ expressed in a system of coordinates $x^a_{-}$, from the exterior spacetime in the $V^{+}$ region, which metric is $g^{+}_{ab}$ expressed in coordinates $x^a_{-}$. If no superscript $\pm$ is given, then the given quantity is considered as a distribution function valid in the entire spacetime. Therefore, the goal is to obtain the necessary condition(s) that must be put on the metric $g_{ab}$ to ensure that $V^+$ and $V^-$ are joined smoothly at $\Sigma$. 

To start this analysis, we must first define some new geometric quantities:
\begin{itemize}
    \item $e^a_\alpha=\frac{\partial x^a}{\partial x^\alpha}$ is the projection tensor from a 4-dimensional spacetime ($V^-$ or $V^+$ region) expressed in $x^a=(t,r,\theta,\varphi)$ coordinates, onto $\Sigma$ with coordinates $x^\alpha=(\tau,\theta,\varphi)$.
    \item $n^a$ is the unit normal vector on $\Sigma$, pointing from $V^-$ to $V^+$, and satisfies the normalization condition $g_{ab}n^a n^b =\epsilon$. This vector is expressed as $n_a=\epsilon \partial_a l$, where $l$ represents the affine parameter along the geodesics that are orthogonal to $\Sigma$, and $\epsilon$ takes on values of $1$, $-1$, or $0$ corresponding to spacelike, timelike, and null geodesic congruences, respectively. Note that by definition the $n^a$ projection onto $\Sigma$ must be zero
    \begin{equation}
        e^a_\alpha n_a = e^\alpha_a n^a= 0
    \end{equation}
    where $e^\alpha_a = \frac{\partial x^\alpha}{\partial x^a}$ is the inverse projection tensor.
    \item The induced metric on the hypersurface $\Sigma$ is $h_{\alpha \beta}=e^a_\alpha e^b_\beta g_{ab}$, induced from the $V^-$ and $V^+$ regions by the corresponding spacetime.
    \item The extrinsic curvature of $\Sigma$ is given by $K_{\alpha \beta}=\nabla_\alpha n_\beta= e^a_\alpha e^b_\beta \nabla_a n_b$, with $K \equiv h^{\alpha \beta} K_{\alpha \beta}$ being the respective trace. 
    \item The energy-momentum tensor of the thin-shell at $\Sigma$ is represented by $S_{\alpha \beta}$.
    \item We define the change of a given quantity $X$ across $\Sigma$ by the jump of its value from the $V^-$ to the $V^+$ region
    \begin{equation}
        [X] \equiv X^{+}|_\Sigma - X^{-}|_\Sigma.
    \end{equation}
    If the quantity $X$ is continuous across $\Sigma$ then $[X]=0$. Note that this implies that $\left[e^a_\alpha\right]=\left[n^a\right]=0$.
\end{itemize}

Now, to derive the junction conditions, we employ the distribution formalism, where any quantity $X$ and its derivative $\nabla_a X$ can be expressed in terms of distribution functions as
\begin{equation}
    X = X^+ \Theta(l) + X^- \Theta(-l),
    \label{X}
\end{equation}
\begin{equation}
    \nabla_a X = \nabla_a X^+ \Theta(l) + \nabla_a X^- \Theta(-l) + \epsilon n_a [X] \delta (l),
    \label{nablaX}
\end{equation}
where $\Theta(l)$ is the Heaviside distribution function defined as $\Theta=0$ for $l<0$, $\Theta(l)=1$ for $l>0$, and $\Theta(l)=1/2$ for $l=0$, and $\delta (l)$ is the Dirac-delta distribution.

To obtain the junction conditions, we must first write all the quantities appearing in the field equation \eqref{Einstein eq} in the distribution formalism. This way, the metric $g_{ab}$ is given as follows
\begin{equation}
    g_{ab} = g^+_{ab} \Theta(l) + g^-_{ab} \Theta(-l).
    \label{gab distribution}
\end{equation}
Consequently, the Christoffel symbols, given in Eq. \eqref{LeviCivita}, require the calculation of the derivatives of the metric, which from Eq. \eqref{nablaX}, these are given as \cite{Rosa:2023guo}
\begin{equation}
    \partial_c g_{ab} = \partial_c g^+_{ab} \Theta(l) + \partial_c g^-_{ab} \Theta(-l) + \epsilon n_c [g_{ab}] \delta(l)
    \label{gab derivative}
\end{equation}
This last term, proportional to $\delta(l)$, poses a challenge. Note that the Christoffel symbols involve products of the inverse metric and its derivatives, consequently containing terms proportional to $\Theta(l) \delta(l)$. The problem arises  when calculating the Riemann tensor, as it involves products of the Christoffel symbols, leading to terms proportional to $\delta^2(l)$ that are singular in the distribution formalism. The junction conditions are precisely introduced to address this problem. In essence, these conditions ensure the elimination of singular terms, particularly those proportional to $\delta^2(l)$, and thus guarantee the well-defined nature of all quantities within the distribution formalism.

Thus, the initial junction condition arises from the requirement to set $[g_{ab}] = 0$, implying that the metric must exhibit continuity across $\Sigma$, in order to eliminate the term proportional to $\delta(l)$ in Eq.\eqref{gab derivative}. Since $[e^a_\alpha]=0$, we can express this condition in a coordinate-independent manner by projecting both indices into the hypersurface $\Sigma$, from which we obtain the following
\begin{equation}
    [h_{\alpha \beta}]=0.
\end{equation}
This is precisely the first junction condition, which requires that the induced metric must be continuous across $\Sigma$.

With this result, Eq. \eqref{gab derivative} simplifies to
\begin{equation}
    \partial_c g_{ab} = \partial_c g^+_{ab} \Theta(l) + \partial_c g^-_{ab} \Theta(-l)
\end{equation}
We are now able to construct the Christoffel symbols in the distribution formalism and, therefore, the Riemann tensor, the Ricci tensor and the Ricci scalar, respectively, which are all regular. These quantities take the following forms \cite{Rosa:2023guo}
\begin{equation}\label{eq:dist_Rabcd}
R_{abcd}=R^+_{abcd}\theta\left(l\right)+R^-_{abcd}\theta\left(-l\right)+\bar R_{abcd}\delta\left(l\right),
\end{equation}
\begin{equation}\label{eq:dist_Rab}
R_{ab}=R^+_{ab}\theta\left(l\right)+R^-_{ab}\theta\left(-l\right)+\bar R_{ab}\delta\left(l\right),
\end{equation}
\begin{equation}\label{eq:dist_R}
R=R^+\theta\left(l\right)+R^-\theta\left(-l\right)+\bar R\delta\left(l\right),
\end{equation}
where the terms $\bar R_{abcd}$,$\bar R_{ab}$ and $\bar R$ denote the terms proportional to $\delta(l)$. In terms of geometrical quantities, these are given by \cite{Rosa:2023guo}
\begin{equation}\label{eq:def_barRabcd}
\bar R_{abcd}=4\left[K_{\alpha\beta}\right]e^\alpha_{[a}n_{b]}e^\beta_{[d}n_{c]},
\end{equation}
\begin{equation}\label{eq:def_barRab}
\bar R_{ab}=-\left(\epsilon\left[K_{\alpha\beta}\right]e^\alpha_a e^\beta_b+n_a n_b \left[K\right]\right),
\end{equation}
\begin{equation}\label{eq:def_barR}
\bar R=-2\epsilon\left[K\right],
\end{equation}
where we have introduced the notation $X_{[ab]}=\frac{1}{2}\left(X_{ab}-X_{ba}\right)$, that denotes the definition of index anti-symmetrization for a given tensor $X_{ab}$. Regarding the field equation \eqref{Einstein eq}, after using the first junction condition to eliminate the singular term, some terms proportional to $\delta(l)$ may still remain in the geometric part of the equation. Thus, it is conceivable that similar terms may appear in the matter part as well. Therefore, it is useful to express the energy-momentum in the distribution formalism as
\begin{equation}\label{eq:dist_tab1}
T_{ab}=T_{ab}^+\theta\left(l\right)+T_{ab}^-\theta\left(-l\right)+S_{ab}\delta\left(l\right),
\end{equation}
where $S_{ab}=S_{\alpha\beta}e^\alpha_a e^\beta_b$. This way, the matching is smooth if $S_{ab}=0$ or there is a shell of matter at $\Sigma$, the so-called thin-shell, when $S_{ab}\neq 0$.  

We now plug the obtained results for $g_{ab}$, $R_{ab}$, $R$ and $T_{ab}$, given in Eqs. \eqref{gab distribution},\eqref{eq:dist_Rab}, \eqref{eq:dist_R} and \eqref{eq:dist_tab1}, respectively, into the field equation \eqref{Einstein eq}. After projecting the result onto $\Sigma$ using $e^a_\alpha e^b_\beta$, and cancelling out any terms not proportional to $\delta(l)$ we obtain
\begin{equation}
    -\epsilon([K_{\alpha \beta}]+h_{\alpha \beta}[K])=\kappa S_{\alpha \beta}.
\end{equation}
From this, we conclude that a smooth matching at $\Sigma$ requires that $[K_{\alpha \beta}]=0$, i.e. the extrinsic curvature must be continuous across the hypersurface. If this condition is violated, then the spacetime is singular at $\Sigma$, as the complete energy-momentum tensor of the hypersurface is $T^{ab}_\Sigma = \delta (l) S^{\alpha \beta} e^a_\alpha e^b_\beta$. However, note that this singularity comes with a physical interpretation: a surface layer with energy-momentum tensor $T^{ab}_\Sigma$ or $S_{\alpha \beta}$ is present at the hypersurface $\Sigma$.

Finally, let us summarize the obtained results. We have found that the junction conditions for a smooth matching of two metrics at the hypersurface $\Sigma$ are given by
\begin{equation}
    [h_{\alpha \beta}]=[K_{\alpha \beta}]=0, 
\end{equation}
indicating that both the induced metric and the extrinsic curvature must be continuous across the hypersurface.

%%%%%%%%%%%%%%%%%%%%%%%%%%%%%%%%%%%%%%%%%%%%%%%%%%%%%%
\section{Wormholes in General Relativity}
%%%%%%%%%%%%%%%%%%%%%%%%%%%%%%%%%%%%%%%%%%%%%%%%%%%%%%

In Chapter \ref{ch:intro}, we highlight the turning point in the physics community to investigate wormholes, occurring when Morris and Thorne \cite{Morris:1988cz} recognized the theoretical feasibility of constructing traversable wormhole spacetimes. The crucial condition for these traversable wormholes is the absence of event horizons and the presence of the throat, or more formally, the geometry must satisfy the flaring-out condition (in subsection \ref{sec:whmetric} we present what this condition entails).

The conventional approach to find solutions of the Einstein equations is to consider a preferred Lagrangian for the matter fields presumed to support a given geometry. Following this, one computes the corresponding energy-momentum tensor and solves the Einstein field equations. The final step involves examining the obtained geometrical solution. Regarding wormholes, Morris and Thorne simplified the analysis by adopting an engineering-oriented perspective. They proposed a well-behaved geometry, in the sense that it does not present singularities. Subsequently, the Riemann tensor associated with this geometry is computed, and the Einstein field equations are applied to deduce the required energy-momentum distribution needed to support this geometry. The final step involves questioning whether the deduced energy-momentum distribution violates any fundamental physical principle.

%%%%%%%%%%%%%%%%%%%%%%%%%%%%%%%%%%%%%%%%%%%%%%%%%%%%%%
\subsection{Wormhole metric and equations}\label{sec:whmetric}
%%%%%%%%%%%%%%%%%%%%%%%%%%%%%%%%%%%%%%%%%%%%%%%%%%%%%%

Let us then introduce their procedure, by first stating the considerations they have made:
\begin{enumerate}
    \item To simplify the calculations, the metric is both spherically symmetric and static (time independent). 
    \item They presuppose the validity of the General Relativity theory, therefore the solutions must obey the Einstein field equations everywhere.
    \item To qualify as a wormhole, the solution must obey the flaring-out condition.
    \item The absence of a horizon is essential, as the presence of one would prevent bidirectional travel through the wormhole.
    \item The tidal gravitational forces experienced by a traveler must be bearably small.
    \item The traveler should traverse the wormhole in a finite and reasonably short proper time (e.g., less than a year), as measured not only by themselves, but also by observers who remain outside the wormhole.
    \item The matter and fields responsible for generating the spacetime curvature of the wormhole must possess a physically reasonable energy-momentum tensor.
\end{enumerate}
In our work, considerations 5 and 6 are not analyzed, therefore we do not present this analysis in General Relativity either. It turns out that the first four considerations restrict the form of the energy-momentum tensor, in a way that near the wormhole's throat it is unusual but does not seem to contradict the known physics. 

Throughout this section, we follow Ref. \cite{Lobo:2007zb} closely. We begin this analysis with the consideration of the following spherically symmetric and static wormhole metric, written in spherical coordinates $x^a={t,r,\theta,\varphi}$
\begin{equation}
    d s^2=-e^{\zeta(r)} d t^2+\frac{d r^2}{1-b(r) / r}+r^2\left(d \theta^2+\sin ^2 \theta d \varphi^2\right),
    \label{metric}
\end{equation}
where $\zeta(r)$ and $b(r)$ are arbitrary functions of the radial coordinate $r$. $\zeta(r)$ is referred to as the redshift function, as it is associated with gravitational redshift, and $b(r)$ is termed the shape function, determining the wormhole's shape through embedding diagrams. The coordinate $r$ is non-monotonic, decreasing from $+\infty$ to a minimum value $r_0$ (representing the throat of the wormhole where $b(r_0) = r_0$), and then increasing from $r_0$ to $+\infty$. For the wormhole to be traversable, it must have no horizons, implying $g_{tt} = -e^{\zeta(r)} \neq 0$. Therefore, $\zeta(r)$ must be finite everywhere.

We can use embedding diagrams, which are graphical representations of a curved surface embedded in a higher-dimensional space, to illustrate and gain valuable insights into the structure of a wormhole. These diagrams help in making informed choices regarding the shape function, $b(r)$. Given the spherical symmetry of the problem, an equatorial slice with $\theta = \pi/2$ can be considered without loss of generality. The corresponding line element for a fixed moment in time, $t = \text{const}$, is given by
\begin{equation}
    d s^2=\frac{d r^2}{1-b(r) / r}+r^2 d \varphi^2.
    \label{slice}
\end{equation}
To visualize this slice, the metric is embedded into three-dimensional Euclidean space, where the metric can be expressed in cylindrical coordinates, $(r, \varphi, z)$ as
\begin{equation}
    d s^2=d z^2+d r^2+r^2 d \varphi^2.
    \label{euclidean metric}
\end{equation}
Now, in three-dimensional Euclidean space, the embedded surface has the equation $z = z(r)$, and thus, the metric of the surface can be expressed as
\begin{equation}
    d s^2=\left[1+\left(\frac{d z}{d r}\right)^2\right] d r^2+r^2 d \varphi^2 .
    \label{euclidean metric 2}
\end{equation}
Comparing Eq. (\ref{euclidean metric 2}) with (\ref{slice}), we obtain the equation for the embedding surface, given by
\begin{equation}
   \frac{dz}{dr}=\pm \left(\frac{r}{b(r)} -1 \right)^{-1/2}.
\end{equation}

\begin{figure}
    \centering
    \includegraphics[scale=0.65]{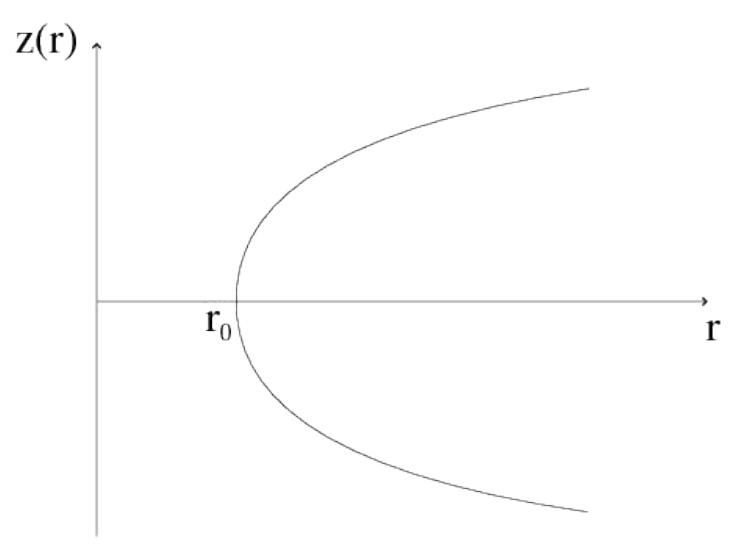}\qquad
    \includegraphics[scale=0.2]{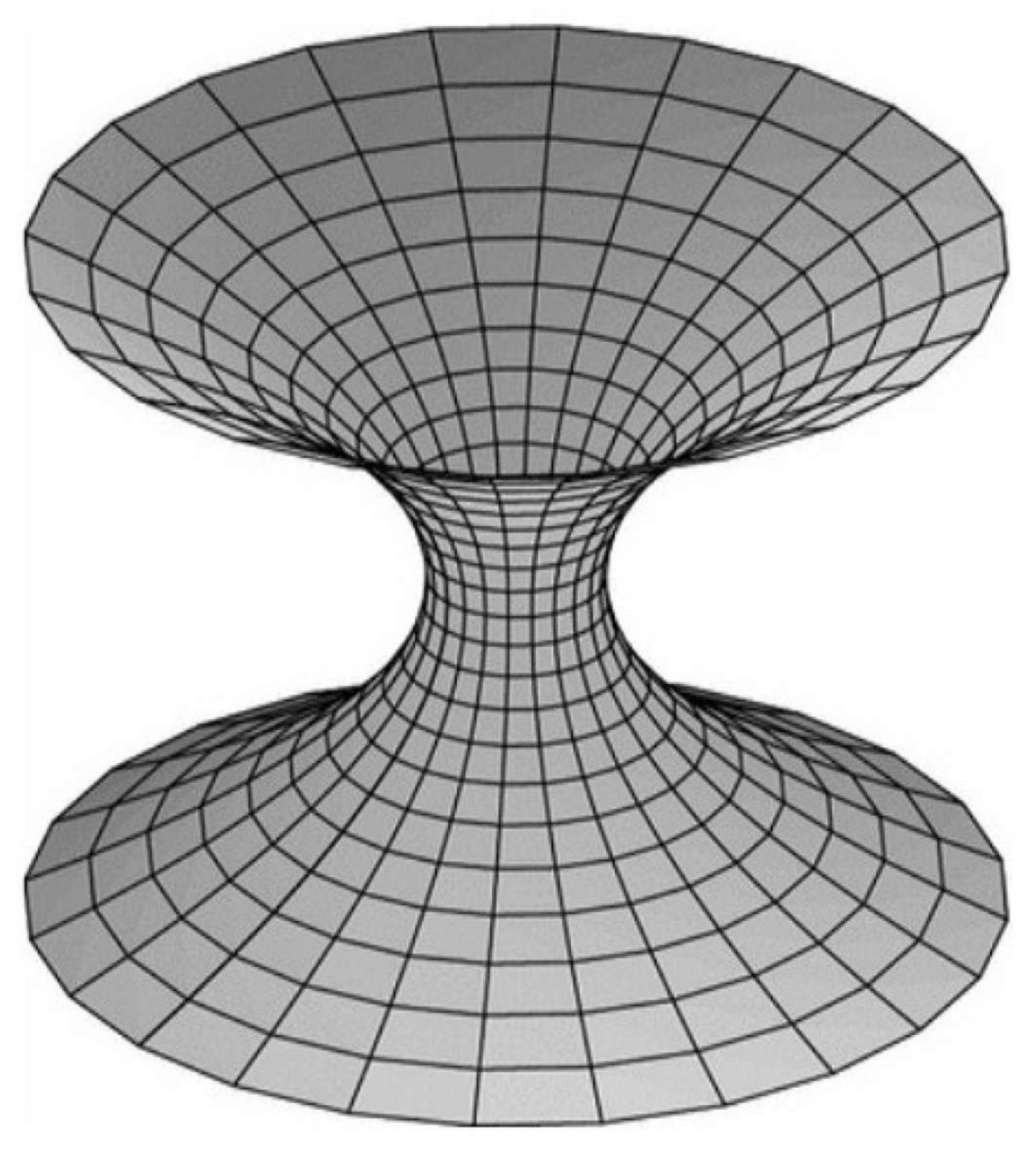}\\
    \caption{Embedding diagram of a two-dimensional slice along the equatorial plane ($t=\text{const},\theta=\pi/2$) of a traversable wormhole (left panel), and illustration of the complete surface sweep through a $2\pi$ rotation around the $z$-axis for a full visualization (right panel) \cite{Lobo:2007zb}.}
    \label{fig:embedding}
\end{figure}

To be considered a wormhole solution, the geometry must feature a minimum radius at the throat, satisfying $b(r_0) = r_0$. At this throat, the embedded surface becomes vertical, indicated by $dz/dr \rightarrow \infty$, as illustrated in Figure \ref{fig:embedding}. Moving away from the throat, the condition is such that space asymptotically tends towards flatness, implying $dz/dr \rightarrow 0$ as $r \rightarrow \infty$.

Additionally, to meet the criteria for a wormhole solution, the throat must demonstrate a flaring-out behavior, as illustrated in Figure \ref{fig:embedding}. Mathematically, this flaring-out condition implies that the second derivative of the inverse of the embedding function, denoted as $r(z)$, must be positive, i.e., $d^2 r/d z^2 > 0$, at or near the throat $r_0$. Upon differentiating the expression $d r / d z = \pm (r / b(r) - 1)^{1 / 2}$ with respect to $z$, we obtain the following
\begin{equation}
    \frac{d^2 r}{d z^2}=\frac{b-b^{\prime} r}{2 b^2}>0.
    \label{flaring-out condition}
\end{equation}
At the throat, we verify that the shape function satisfies the condition $b^\prime (r_0)<1$. It will become evident in the subsequent discussion that this condition plays a pivotal role in the analysis of the violation of the energy conditions.

To solve the Einstein field equations, we first note that the mathematical analysis and the physical interpretation will be simplified if one employs a set of orthonormal basis vectors. These vectors can be interpreted as the proper reference frame of observers who remain stationary in the coordinate system $(t, r, \theta, \varphi)$, with $(r, \theta, \varphi)$ held constant. Let us denote the basis vectors in the coordinate system as $\left(\mathbf{e}_t, \mathbf{e}_r, \mathbf{e}_\theta, \mathbf{e}_\varphi\right)$. Thus, the orthonormal basis vectors take the form
\begin{equation}
    \left\{\begin{array}{l}
\mathbf{e}_{\hat{t}}=e^{-\zeta /2} \mathbf{e}_t ,\\
\mathbf{e}_{\hat{r}}=(1-b / r)^{1 / 2} \mathbf{e}_r ,\\
\mathbf{e}_{\hat{\theta}}=r^{-1} \mathbf{e}_\theta, \\
\mathbf{e}_{\hat{\varphi}}=(r \sin \theta)^{-1} \mathbf{e}_\varphi .
\end{array}\right .
\end{equation}
The Einstein tensor, expressed in the orthonormal reference frame as $G_{\hat{a} \hat{b}}=R_{\hat{a} \hat{b}}-\frac{1}{2} R g_{\hat{a} \hat{b}}$, produces the following non-zero components for the metric (\ref{metric})
\begin{equation}
G_{\hat{t} \hat{t}} =\frac{b^{\prime}}{r^2} ,
\label{Gtt}
\end{equation}
\begin{equation}
G_{\hat{r} \hat{r}}  =-\frac{b}{r^3}+2\left(1-\frac{b}{r}\right) \frac{\zeta^{\prime}}{2r} ,
\end{equation}
\begin{equation}
G_{\hat{\theta} \hat{\theta}} =\left(1-\frac{b}{r}\right)\left[\frac{\zeta^{\prime \prime}}{2}+\left(\frac{\zeta^{\prime}}{2}\right)^2-\frac{b^{\prime} r-b}{2 r(r-b)} \zeta^{\prime}-\frac{b^{\prime} r-b}{2 r^2(r-b)}+\frac{\zeta^{\prime}}{r}\right] ,
\end{equation}
\begin{equation}
G_{\hat{\varphi} \hat{\varphi}}  =G_{\hat{\theta} \hat{\theta}} .
\label{Gphiphi}
\end{equation}
The Einstein field equation, given by $G_{\hat{a} \hat{b}}=8 \pi T_{\hat{a} \hat{b}}$, imposes that the energy-momentum tensor $T_{\hat{a} \hat{b}}$ should have the same algebraic structure as the Einstein tensor, given in Eqs. (\ref{Gtt})-(\ref{Gphiphi}). Therefore, the nonzero components are precisely the diagonal terms $T_{\hat{t} \hat{t}},T_{\hat{r} \hat{r}},T_{\hat{\theta} \hat{\theta}},T_{\hat{\varphi} \hat{\varphi}}$. Using the orthonormal basis, these components carry a simple physical interpretation, i.e., we have that 
\begin{equation}
    T_{\hat{t} \hat{t}}=\rho(r), \quad T_{\hat{r} \hat{r}}=p_r(r), \quad T_{\hat{\theta} \hat{\theta}}=T_{\hat{\varphi} \hat{\varphi}}=p_t(r).
\end{equation}
In this expression, $\rho(r)$ represents the energy density, $p_r(r)$ denotes the radial pressure, and $p_t(r)$ is the pressure measured in the tangential directions, orthogonal to the radial direction. We can also denote the radial pressure in terms of the radial tension, $\tau(r)$, knowing that $p_r(r)=-\tau(r)$. Using the Einstein field equation, $G_{\hat{a}\hat{b}}=8\pi T_{\hat{a}\hat{b}}$, we obtain the following energy-momentum expressions
\begin{equation}
    \rho(r)=\frac{1}{8 \pi} \frac{b^{\prime}}{r^2} ,
    \label{rho1}
\end{equation}
\begin{equation}
    p_r(r)=\frac{-1}{8 \pi}\left[\frac{b}{r^3}-2\left(1-\frac{b}{r}\right) \frac{\zeta^{\prime}}{2r}\right] ,
    \label{pr1}
\end{equation}
\begin{equation}
    p_t(r)=\frac{1}{8 \pi}\left(1-\frac{b}{r}\right)\left[\frac{\zeta^{\prime \prime}}{2}+\left(\frac{\zeta^{\prime}}{2}\right)^2-\frac{b^{\prime} r-b}{2 r(r-b)} \zeta^{\prime}-\frac{b^{\prime} r-b}{2 r^2(r-b)}+\frac{\zeta^{\prime}}{r}\right] .
    \label{pt1}
\end{equation}

%%%%%%%%%%%%%%%%%%%%%%%%%%%%%%%%%%%%%%%%%%%%%%%%%%%%%%
\subsection{Energy conditions}
%%%%%%%%%%%%%%%%%%%%%%%%%%%%%%%%%%%%%%%%%%%%%%%%%%%%%%

To better understand the nature of the matter within the wormhole, Morris and Thorne \cite{Morris:1988cz} introduced the dimensionless function $\xi=(\tau-\rho) /|\rho|$. Using equations (\ref{rho1})-(\ref{pr1}), knowing that $p_r(r)=-\tau(r)$, one obtains
\begin{equation}
    \xi=\frac{\tau-\rho}{|\rho|}=\frac{b / r-b^{\prime}-r(1-b / r) \zeta^{\prime}}{\left|b^{\prime}\right|}.
    \label{exoticity}
\end{equation}
Combining Eq. (\ref{exoticity}) with the flaring-out condition, Eq. (\ref{flaring-out condition}), the exoticity function, $\xi(r)$, takes the form
\begin{equation}
    \xi=\frac{2 b^2}{r\left|b^{\prime}\right|} \frac{d^2 r}{d z^2}- r\left(1-\frac{b}{r}\right) \frac{\zeta^{\prime}}{\left|b^{\prime}\right|}.
\end{equation}
Given the finite nature of $\rho$ and, consequently, $b^\prime$, along with the fact that $(1-b/r) \zeta^\prime \rightarrow 0$ at the throat, we establish the following relationship
\begin{equation}
    \xi\left(r_0\right)=\frac{\tau_0-\rho_0}{\left|\rho_0\right|}>0 .
\end{equation}
The constraint $\tau_0 > \rho_0$ poses a significant challenge, as it implies that the radial tension at the throat must exceed the energy density. Consequently, Morris and Thorne referred to matter satisfying this condition as "exotic matter." Subsequent examination will confirm that this type of matter violates the null energy condition (in fact, it violates all the energy conditions).

Let us then delineate the energy conditions, focusing on the scenario where the energy-momentum tensor is diagonal, i.e.
\begin{equation}
    T^a_b=\operatorname{diag}\left(-\rho, p_1, p_2, p_3\right).
    \label{diagT}
\end{equation}
Here, $\rho$ denotes the energy density, and $p_j$ represents the three principal pressures. When $p_1=p_2=p_3$, this formulation reduces to the energy-momentum tensor of a perfect fluid. Although classical forms of matter are generally expected to obey to these energy conditions, it is worth noting that they can be violated by certain quantum fields, with the Casimir effect being one notable example.

\textbf{Null energy condition (NEC)}. The NEC asserts that for any null vector $k^a$ 
\begin{equation}
    T_{ab} k^a k^b \geq 0.
\end{equation}
In the case of a energy-momentum tensor of the form (\ref{diagT}), we have 
\begin{equation}
    \forall i, \quad \rho+p_i \geq 0.
\end{equation}
We can interpret $T_{ab} k^a k^b$ as the energy density measured by any null observer with four-velocity $k^a$.

\textbf{Weak energy condition (WEC)}. The WEC states that for any timelike vector $U^a$
\begin{equation}
    T_{ab} U^a U^b \geq 0.
\end{equation}
We can interpret $T_{ab} U^a U^b$ as the energy density measured by any timelike observer with four-velocity $U^a$. Therefore, the WEC stipulates that this quantity must be positive. Expressed in terms of the principal pressures, this condition can be stated as
\begin{equation}
    \rho \geq 0  \quad \text{ and } \quad \forall i, \quad \rho+p_i \geq 0.
\end{equation}
By continuity, the WEC implies the NEC.

\textbf{Strong energy condition (SEC)}. The SEC asserts that for any timelike vector $U^a$ the following inequality holds
\begin{equation}
    \left(T_{ab}-\frac{T}{2} g_{ab}\right) U^a U^b \geq 0,
\end{equation}
where $T=g^{ab}T_{ab}$ is the trace of the energy-momentum tensor. Considering the diagonal energy-momentum tensor (\ref{diagT}), the SEC can be expressed as
\begin{equation}
    \forall i, \quad \rho+p_i \geq 0 \quad \text { and } \quad \rho+\sum_i p_i \geq 0 .
\end{equation}
The SEC, linked to preserving the attractive character of gravity of the object under study, implies the NEC but not necessarily the WEC.

\textbf{Dominant energy condition (DEC)}. The DEC states that for any timelike vector $U^a$
\begin{equation}
    T_{ab} U^a U^b \geq 0 \quad \text{and} \quad  T_{ab} U^b \quad \text{is not spacelike}
\end{equation}
These conditions indicate that the locally observed energy density should be positive, and the energy flux should be either timelike or null. Note that the DEC implies the WEC and, consequently, the NEC, but not necessarily the SEC. For a energy-momentum tensor in the form of (\ref{diagT}), we find
\begin{equation}
    \rho \geq 0 \quad \text{and} \quad \forall i, \quad p_i \in [-\rho,+\rho].
\end{equation}

This condition ensures that the speed of sound is smaller than the speed of light $c$, and is frequently associated with the stability of the object under study.

In summary, the analysis focuses on four distinct energy conditions: NEC, WEC, SEC, and DEC. When considering a diagonal energy-momentum tensor $T_{ab}$ in the form of Eq. (\ref{diagT}), these energy conditions take the following forms:
\begin{align}
    \text{NEC:} & \quad \rho+p_r \geq 0,\quad \rho+p_t \geq 0, \label{NEC}\\
    \text{WEC:} & \quad \text{NEC and } \rho \geq 0, \label{WEC}\\
    \text{SEC:} & \quad \text{NEC and } \rho+p_r+2p_t \geq 0, \label{SEC}\\
    \text{DEC:} & \quad \rho \geq |p_r|, \quad \rho \geq |p_t|. \label{DEC}
\end{align}

Now, applying the Einstein field equations (\ref{rho1}) and (\ref{pr1}) at the throat $r_0$ and considering the finite nature of the redshift function, the flaring-out condition imposes that  $\left.\left(\rho+p_r\right)\right|_{r_0}<0$. This violation contradicts the NEC, indicating a breach of all pointwise energy conditions. While classical forms of matter are expected to satisfy the energy conditions, it is noteworthy that certain quantum fields, like in the case of the Casimir effect, are known to violate them.

Hence, the flaring-out condition (\ref{flaring-out condition}) implies the violation of the NEC at the throat. In other words, it is the geometry of the wormhole that causes the NEC violation, suggesting that it must be sustained by exotic matter. Note that negative energy densities are not strictly required; however, negative pressures at the throat, such as $p_r(r_0)=-1/(8\pi r_0^2)$, are essential to maintain the wormhole throat. Looking at it classically, challenging the null energy condition seems questionable at most. However, we do know that certain quantum effects lead to small, experimentally verified, violations of the null energy condition. Without the experimental evidence indicating that quantum effects violate this condition, Morris and Thorne's work might have been perceived as the initial steps towards formulating a theorem asserting the impossibility of wormholes. Conversely, it is crucial to acknowledge that the observed violations of the null energy condition are minimal, if not infinitesimal. Whether it is possible to achieve a violation substantial enough to sustain a traversable wormhole remains far from certain.

Recognizing this challenge, efforts have been made to minimize this issue. Various strategies extensively explored in the literature include rotating solutions \cite{Teo:1998dp}, evolving wormhole spacetimes \cite{Kar:1994tz,Kar:1995ss,Arellano:2006ex}, thin-shell wormholes employing the cut-and-paste procedure \cite{Poisson:1995sv}-\cite{Garcia:2011aa}, and modifications to theories of gravity \cite{Capozziello:2012hr}-\cite{Harko:2013yb} (see Chapter \ref{ch:intro} for more details).

%%%%%%%%%%%%%%%%%%%%%%%%%%%%%%%%%%%%%%%%%%%%%%%%%%%%%%
\chapter{$f(R,\mathcal{T})$ modified-gravity theory}\label{ch: MGT}\label{ch:modifiedgravity}
%%%%%%%%%%%%%%%%%%%%%%%%%%%%%%%%%%%%%%%%%%%%%%%%%%%%%%

To overcome the reliance on the exotic matter, in this chapter we analyze wormhole spacetimes within the framework of modified theories of gravity. In these modified theories, the field equations can be reformulated as an effective Einstein equation \cite{Harko:2013yb} 
\begin{equation}
    G_{ab} = \kappa T_{ab}^{\text{eff}},
\end{equation}
where $T_{ab}^{\text{eff}}$ is the effective energy-momentum tensor that contains both the matter energy-momentum tensor $T_{ab}$ as well as the curvature quantities specific to the modified theory of gravity under consideration. This way, the generalized null energy condition is modified to the following \cite{Capozziello:2013vna,Capozziello:2014bqa}
\begin{equation}
    T_{ab}^{\text{eff}} k^a k^b \geq 0.
    \label{generalized NEC}
\end{equation}
Therefore, by definition, the necessary condition for the existence of the wormhole geometry is the violation of the generalized NEC Eq.\eqref{generalized NEC}, as it will ensure that the flaring-out condition Eq. \eqref{flaring-out condition} is satisfied. Now, in Chapter \ref{ch:framework}, in the context of General Relativity theory, this condition simplifies to the violation of the standard NEC, $T_{ab} k^a k^b <0$, which implies the use of the exotic matter to sustain these geometries. However, in the case of modified theories of gravity, one may, in principle, build a theory that enforces the matter energy-momentum tensor to satisfy the standard NEC, $T_{ab} k^a k^b \geq 0$, while the generalized NEC is necessarily violated $T_{ab}^{\text{eff}} k^a k^b <0$. This way, we have simultaneously the fulfillment of the flaring-out condition, ensuring the wormhole geometry, as well as the presence of normal matter, avoiding completely the use of the exotic one. 

%%%%%%%%%%%%%%%%%%%%%%%%%%%%%%%%%%%%%%%%%%%%%%%%%%%%%%
\section{Theory and framework}\label{sec:theory}
%%%%%%%%%%%%%%%%%%%%%%%%%%%%%%%%%%%%%%%%%%%%%%%%%%%%%%

\subsection{Action and field equations of $f\left(R,\mathcal T\right)$}

In this work, the chosen theory is a covariant generalization of Einstein's General Relativity known as energy-momentum squared gravity, or $f\left(R,\mathcal T\right)$ gravity \cite{katirci1,roshan1}, where $R$ is the Ricci scalar and $\mathcal T=T_{ab}T^{ab}$ with $T_{ab}$ being the energy-momentum tensor. Therefore, the action that describes this theory is similar to the Einstein-Hilbert's, where we replace $R$ by $f(R,\mathcal{T})$, that is
\begin{equation}\label{geo_action}
S=\frac{1}{2\kappa^2}\int_\Omega \sqrt{-g}f\left(R,\mathcal T\right)d^4x+\int_\Omega \sqrt{-g}\mathcal L_m d^4x,
\end{equation}
where $\Omega$ is a spacetime manifold with coordinates $x^a$. 

As previously, the energy-momentum tensor is obtained through the variation of the matter Lagrangian $\mathcal L_m$ with respect to the metric, resulting in
\begin{equation}\label{def_tab}
T_{ab}=-\frac{2}{\sqrt{-g}}\frac{\delta\left(\sqrt{-g}\mathcal L_m\right)}{\delta g^{ab}}.
\end{equation}
Considering that the matter Lagrangian only depends on the metric components and not on their derivatives, this expression becomes \cite{Katirci:2013okf}
\begin{equation}
    T_{ab} = g_{ab} \mathcal{L}_m - 2\frac{\partial \mathcal{L}_m}{\partial g^{ab}}.
\end{equation}
Now, using the variational method on Eq.\eqref{geo_action} with respect to the metric $g_{ab}$ results in the modified field equations, given by
\begin{equation}\label{geo_field}
f_R R_{ab}-\frac{1}{2}g_{ab}f-\left(\nabla_a\nabla_b-g_{ab}\Box\right)f_R=8\pi T_{ab}-f_\mathcal T \Theta_{ab},
\end{equation}
where $f_R\equiv \partial f/\partial R$ and $f_\mathcal T=\partial f/\partial \mathcal T$ and $\Box=g^{ab}\nabla_a\nabla_b$ is the d'Alembert operator. Additionally, $\Theta_{ab}$ is an auxiliary quantity arising from the variation of $\mathcal{T}$ as follows
\begin{equation}
\Theta_{ab}=\frac{\delta\mathcal T}{\delta g^{ab}},
\end{equation}
which becomes \cite{Katirci:2013okf}
\begin{equation}
    \Theta_{ab} = -2\mathcal{L}_m\left(T_{ab} - \frac{1}{2}g_{ab}T\right)-T T_{ab} + 2T^c_a T_{bc} - 4T^{cd}\frac{\partial^2\mathcal{L}_m}{\partial g^{ab} \partial g^{cd}}.
\end{equation}
The explicit form of this auxiliary tensor is obtained once we specify a choice for the matter Lagrangian $\mathcal L_m$ or, analogously, a choice for the energy-momentum tensor. Regarding Eq. \eqref{geo_field}, we can take the covariant derivative of it in order to obtain the following conservation equation
\begin{equation}\label{geo_conservation}
8\pi \nabla_bT^{ab}=\nabla_b\left(f_\mathcal T \Theta^{ab}\right)+f_R\nabla_bR^{ab}-\frac{1}{2}g^{ab}\nabla_bf.
\end{equation}
Note that this result is different from GR, where in general one obtains $\nabla_b T^{ab}=0$, while in this case we see that Eq. \eqref{geo_conservation} does not imply the conservation of $T_{ab}$. This suggests a flow of energy between the matter sector and the gravitational one, and vice versa, allowing for instance the possibility of gravitationally induced matter creation (see Ref. \cite{Cipriano:2023yhv} for more details).

Furthermore, for simplicity, we consider that the function $f\left(R,\mathcal T\right)$ is separable and linear in both $R$ and $\mathcal T$
\begin{equation}
    f\left(R,\mathcal T\right)=R+\gamma \mathcal T,
    \label{linear_f}
\end{equation}
where $\gamma$ is a coupling constant. This way, Eq. \eqref{geo_field} and the conservation equation simplifies to \eqref{geo_conservation}
\begin{equation}\label{field}
G_{ab}=8\pi T_{ab}-\gamma\left(\Theta_{ab}-\frac{1}{2}g_{ab}\mathcal T\right),
\end{equation}
\begin{equation}\label{conservation}
8\pi \nabla_b T^{ab}=\gamma\nabla_b\left(\Theta^{ab}-\frac{1}{2}g^{ab}\mathcal T\right),
\end{equation}
respectively. Later, in section \ref{sec:extensions} we extend this assumption to the case of higher powers of the term $\mathcal{T}$.  

\subsection{Wormhole solutions in $f(R,\mathcal{T})=R+\gamma \mathcal{T}$ gravity}\label{subsec:worm sol}

Similar to Morris and Thorne's considerations, in this work we are interested in static and spherically symmetric traversable wormhole solutions, which metric is given by Eq.(\ref{metric}). For the wormhole to be traversable, let us remember that the functions $\zeta (r)$ and $b(r)$ must satisfy a few conditions. First, in order to avoid event horizons the redshift function must remain finite throughout the whole spacetime, i.e., $|\zeta (r)|< \infty$. Additionally, from the flaring-out condition at the throat, the shape function must obey $b\left(r_0\right)=r_0$ and $ b'\left(r_0\right)<1$. The following two broad families of solutions for the functions $\zeta (r)$ and $b(r)$ satisfy the requirements above
\begin{equation}\label{zbfunctions}
\zeta\left(r\right)=\zeta_0\left(\frac{r_0}{r}\right)^\alpha, \qquad b\left(r\right)=r_0\left(\frac{r_0}{r}\right)^\beta,
\end{equation}
where $\zeta_0$ is an arbitrary constant, and $\alpha$ and $\beta$ are arbitrary positive exponents. 

Now, regarding the matter sector, we assume that the energy-momentum tensor takes the form of a distribution of matter that is well described by an anisotropic perfect fluid
\begin{equation}\label{def_matter}
T_a^b=\text{diag}\left(-\rho,p_r,p_t,p_t\right).
\end{equation}
In fact, $\rho=\rho(r)$, $p_r=p_r(r)$ and $p_t=p_t(r)$, i.e., these quantities are assumed to depend only on the radial coordinate $r$ in order to preserve the spherical symmetry of the wormhole. This way, the matter Lagrangian $\mathcal{L}_m$ takes the form
\begin{equation}
    \mathcal{L}_m = \frac{1}{3} (p_r + 2 p_t),
\end{equation}
from which we compute the following auxiliary tensor $\Theta_{ab}$
\begin{equation}\label{def_theta_2}
\Theta_{ab}=-\frac{2}{3}\left(p_r+2p_t\right)\left(T_{ab}-\frac{1}{2}g_{ab} T\right)-T T_{ab}+2T_{a}^cT_{cb}.
\end{equation}
Therefore, under these assumptions, the field equations in Eq.\eqref{field} result in
\begin{eqnarray}\label{eqrho}
8\pi\rho &=& \frac{\gamma}{6}\left(p_r^2-2p_t^2-3\rho^2-8p_rp_t-8p_r\rho-16p_t\rho\right)-\frac{\beta}{r^2}\left(\frac{r_0}{r}\right)^{\beta+1},
\end{eqnarray} 
\begin{eqnarray}\label{eqpr}
8\pi p_r &=& \frac{\gamma}{6}\left(p_r^2+2p_t^2-3\rho^2-12p_rp_t+4p_r\rho-4p_t\rho\right)
	\nonumber \\
&& \hspace{-0.5cm} -\frac{1}{r^2}\left(\frac{r_0}{r}\right)^{\beta+1}-\frac{\alpha\zeta_0}{r^2}\left(\frac{r_0}{r}\right)^\alpha\left[1-\left(\frac{r_0}{r}\right)^{\beta+1}\right],
\end{eqnarray}
\begin{eqnarray}\label{eqpt}
8\pi p_t &=&-\frac{\gamma}{6}\left(p_r^2+6p_t^2+3\rho^2+2p_rp_t+2p_r\rho-2p_t\rho\right) +\frac{1+\beta}{2r^2}\left(\frac{r_0}{r}\right)^{\beta+1}
\nonumber \\
&&+\frac{\alpha^2\zeta_0^2}{4r^2}\left(\frac{r_0}{r}\right)^{2\alpha}\left[1-\left(\frac{r_0}{r}\right)^{\beta+1}\right]	+\frac{\alpha\zeta_0}{4r^2}\left(\frac{r_0}{r}\right)^\alpha\left[2\alpha-\left(1+2\alpha+\beta\right)\left(\frac{r_0}{r}\right)^{\beta+1}\right].
\end{eqnarray}
This way, equations \eqref{eqrho}--\eqref{eqpt} form a system of three equations for the three unknowns $\rho$, $p_r$, and $p_t$. Note that these three equations are quadratic in their respective unknowns, indicating that this system yields a maximum of eight independent solutions, some of which may be complex for some specific combinations of parameters. 

However, the system of Eqs.\eqref{eqrho}--\eqref{eqpt} is too complicated to obtain explicit analytical solutions for $\rho$, $p_r$ and $p_t$, even when we consider specific choices for the free parameters $r_0$, $\alpha$, $\beta$, $\gamma$, and $\zeta_0$. Nevertheless, we can still obtain analytical solutions for these quantities in a recursive manner, stated as follows
\begin{itemize}
    \item Choose specific values for the free parameters $r_0$, $\alpha$, $\beta$, $\gamma$, and $\zeta_0$.
    \item For this particular choice, starting at $r=r_0$, solve algebraically the system for $\rho\left(r_0\right)$, $p_r\left(r_0\right)$, and $p_t\left(r_0\right)$, in order to obtain the numerical values of these quantities. This results in a set of values $\left\{\rho_0^i,p_{r0}^i,p_{t0}^i\right\}$, for $i\in\left\{1,...,8\right\}$ corresponding to the eight independent solutions of the system. 
    \item For each of these solutions, increment the radius $r$ in small steps, such as $r_{n+1} = r_{n} + \epsilon$ for some small value $\epsilon$, and find the values of $\rho\left(r_{n+1}\right)$, $p_r\left(r_{n+1}\right)$, and $p_t\left(r_{n+1}\right)$.
    \item Repeat the process up to a radius $r$ large enough.
\end{itemize}
This way, one is able to extract analytically the behaviour of the solutions. Remember that among these obtained solutions, our focus relies on those relevant from an astrophysical standpoint, i.e., those whose matter components obey the energy conditions Eqs.\eqref{NEC}-\eqref{DEC}. In other words, from the set of eight solutions for the matter quantities, we discard those which violate any of the energy conditions stated above. 

%%%%%%%%%%%%%%%%%%%%%%%%%%%%%%%%%%%%%%%%%%%%%%%%%%%%%%
\section{Junction conditions and matching}\label{sec:matching}
%%%%%%%%%%%%%%%%%%%%%%%%%%%%%%%%%%%%%%%%%%%%%%%%%%%%%%

As previously stated, the pursuit of physically relevant spacetime solutions describing localized objects may need the use of junction conditions. In fact, the solutions obtained and presented in Chapter \ref{ch:results} require this consideration, since the matter components do not vanish across the entire range of the radial coordinate, indicating that these are not localized. Therefore, to obtain localized solutions, we must consider a matching at some finite radius with an exterior vacuum spacetime. Since the junction conditions are theory dependent, let us start by deriving them for the $f(R,\mathcal{T})$ linear case and later proceed with the matching of spacetimes. 

\subsection{Junction conditions of linear $f\left(R,\mathcal T\right)$}

To obtain the junction conditions for the linear $f\left(R,\mathcal{T}\right)$ gravity, we use the geometrical quantities and the formalism developed in the subsection \ref{subsec:JC}. Note that this derivation looks the same as in GR, only differing when we introduce the matter sector. Therefore, the first junction condition is the same as previously obtained for GR, that states that the induced metric $h_{ab}$ must be continuous across the hypersurface $\Sigma$, i.e. $[h_{ab}]=0$. 

Considering now the matter sector, let us remember that the goal of the junction conditions is to eliminate singular terms, in particular those proportional to $\delta^2(l)$, to guarantee well-defined quantities in the field equations, which in this case is the modified field equation \eqref{field}. Motivated by this, as previously done in GR, we define the energy-momentum tensor in the distribution formalism as 
\begin{equation}\label{eq:dist_tab}
T_{ab}=T_{ab}^+\theta\left(l\right)+T_{ab}^-\theta\left(-l\right)+S_{ab}\delta\left(l\right),
\end{equation}
with $S_{ab}=S_{\alpha\beta}e^\alpha_a e^\beta_b$, and $S_{\alpha\beta}$ representing the 3-dimensional energy-momentum tensor of the thin-shell. Additionally, in this theory we must define the scalar $\mathcal T$ in the distributional formalism, which can be obtained through the contraction of $T_{ab}$ with itself, taking the following form
\begin{equation}
\mathcal T=\mathcal T^+\theta\left(l\right)+\mathcal T^-\theta\left(-l\right)+\bar{\mathcal T}\delta\left(l\right)+\hat{\mathcal T}\delta^2\left(l\right),
\label{mathcalT_dist}
\end{equation}
where the quantities $\bar{\mathcal T}$ and $\hat{\mathcal T}$ are expressed in terms of matter quantities as
\begin{equation}
\bar{\mathcal T}=\left(T_{ab}^++T_{ab}^-\right)S^{ab},
\end{equation}
\begin{equation}
\hat{\mathcal T}=S_{ab}S^{ab}.
\end{equation}
Note that the term proportional to $\delta^2\left(l\right)$ in Eq.\eqref{mathcalT_dist} is singular in the distributional formalism and needs to be eliminated. However, $\hat{\mathcal T}$ is proportional to a quadratic term in $S_{ab}$, which is then always non-negative. Therefore, the only possible way to eliminate the term proportional to $\delta^2\left(l\right)$ is to force the energy-momentum tensor of the thin-shell to vanish.
\begin{equation}\label{junctionS}
    S_{ab}=0.
\end{equation}
This way, in $f\left(R,\mathcal T\right)$ gravity theory the smooth matching is the only allowed method of matching two spacetimes preserving the regularity of the action, contrary to the General Relativity case where we could choose if the matching was a smooth one or it presented a thin-shell at the hypersurface $\Sigma$.

Following the same procedure as in GR, we now take the definitions outlined above with the restriction that $S_{ab}=0$, to compute the field equations \eqref{field} projected into the hypersurface $\Sigma$ using $e^a_\alpha e^b_\beta$, which take the form
\begin{equation}
    \left[K_{\alpha\beta}\right]-\left[K\right]h_{\alpha\beta}=0.
    \label{quase2JC}
\end{equation}
We can take the trace of this result by multiplying the equation with $h^{\alpha\beta}$, which implies that $\left[K\right]=0$. Replacing this result, $\left[K\right]=0$, back into the Eq.\eqref{quase2JC} leads to
\begin{equation}
\left[K_{\alpha\beta}\right]=0.
\end{equation}
Finally, we have obtained the second junction condition that implies that the extrinsic curvature $K_{\alpha\beta}$ must be continuous across the hypersurface $\Sigma$.

Let us summarize the obtained results. We have shown that the matching between two spacetimes in linear $f\left(R,\mathcal T\right)$ gravity must always be smooth, which means that there is not a thin-shell at $\Sigma$, and the two junction conditions that the spacetimes must satisfy are the same as in GR. These are, the induced metric $h_{\alpha\beta}$ and the extrinsic curvature $K_{\alpha\beta}$ must be continuous across the hypersurface $\Sigma$,
\begin{equation}\label{junction}
\left[h_{\alpha\beta}\right]=0, \qquad \left[K_{\alpha\beta}\right]=0.
\end{equation}

\subsection{Matching with an exterior vacuum}

Let us now perform the matching between the interior wormhole spacetime with an exterior spherically symmetric and static vacuum solution, using the previously derived junction conditions. The metrics of the interior and exterior spacetimes to be matched are 
\begin{equation}\label{metrici}
ds_-^2=-C e^{\zeta_0\left(\frac{r_0}{r}\right)^\alpha}dt^2+\left[1-\left(\frac{r_0}{r}\right)^{\beta+1}\right]^{-1}dr^2+r^2d\Omega^2,
\end{equation}
\begin{equation}\label{metrice}
ds_+^2=-\left(1-\frac{2M}{r}\right)dt^2+\left(1-\frac{2M}{r}\right)^{-1}dr^2+r^2d\Omega^2,
\end{equation}
respectively. Note that the metric in Eq.\eqref{metrici} corresponds to the metric in Eq.\eqref{metric}, where we have introduced the choices \eqref{zbfunctions} for the redshift and shape functions, and the metric in Eq.\eqref{metrice} is the Schwarzschild solution with a mass $M$. Additionally, we have introduced the constant $C$ in Eq.\eqref{metrici} to later ensure that the time coordinates in both the interior and exterior metrics coincide.  

This analysis turns out to be more clear if we start with the second junction condition. Considering the spherical symmetry of both metrics, the extrinsic curvatures $K_{\alpha\beta}^\pm$ feature only two independent components, the $K_{00}$ and $K_{\theta\theta}=K_{\phi\phi}\sin^2\theta$. Explicitly, these components are
\begin{equation}
K_{00}^-=\frac{\alpha\zeta_0}{2r}\left(\frac{r_0}{r}\right)^\alpha\sqrt{1-\left(\frac{r_0}{r}\right)^{\beta+1}},
\end{equation}
\begin{equation}
K_{00}^+=-\frac{M}{r^2}\sqrt{\frac{r}{r-2M}},
\end{equation}
\begin{equation}
K_{\theta\theta}^-=r\sqrt{1-\left(\frac{r_0}{r}\right)^{\beta+1}},
\end{equation}
\begin{equation}
K_{\theta\theta}^+=r\sqrt{1-\frac{2M}{r}}.
\end{equation}
This way, the second junction condition in Eq. \eqref{junction}, i.e., $\left[K_{\alpha\beta}\right]=0$, implies two independent constrains to the matching, which are $\left[K_{00}\right]=0$ and $\left[K_{\theta\theta}\right]=0$. Explicitly, these constrains are given as
\begin{equation}\label{cond1}
\frac{\alpha\zeta_0}{2}\left(\frac{r_0}{r}\right)^\alpha\sqrt{1-\left(\frac{r_0}{r}\right)^{\beta+1}}+\frac{M}{r}\sqrt{\frac{r}{r-2M}}=0,
\end{equation}
\begin{equation}\label{cond2}
\sqrt{1-\left(\frac{r_0}{r}\right)^{\beta+1}}=\sqrt{1-\frac{2M}{r}},
\end{equation}
respectively. To solve these equations, let us begin with Eq.\eqref{cond2} which can be solved for the radius $r=r_\Sigma$ where the matching must occur. It has a unique real solution for $M>0$ and $r_0>0$, which is expressed as follows
\begin{equation}\label{rsigma}
r_\Sigma=\left(2M\right)^{-\frac{1}{\beta}}\left(r_0\right)^{1+\frac{1}{\beta}}.
\end{equation}
Keep in mind that this radius $r_\Sigma$ must satisfy the condition $r_\Sigma>2M$ to prevent the existence of event horizons in the complete wormhole spacetime solution. Revisiting the derived result in Eq.\eqref{rsigma}, this implies that $r_0>2M$ for any $\beta\geq1$. Now, we can substitute the obtained solution for $r_\Sigma$ back into the first condition Eq.\eqref{cond1}. Solving this equation provides the value of $\zeta_0$ for which the matching at the radius $r=r_\Sigma$ is possible. This way, we obtain the following solution for $\zeta_0$
\begin{equation}\label{zsigma}
\zeta_0=\frac{\left(2M\right)^{\frac{1-\alpha+\beta}{\beta}}\left(r_0\right)^{\frac{\alpha}{\beta}}}{\alpha\left[\left(2M\right)^{1+\frac{1}{\beta}}-\left(r_0\right)^{1+\frac{1}{\beta}}\right]}.
\end{equation}
Once again, we need to take into account the condition $r_0>2M$, which, in this case, implies that for any $\alpha\geq 1$ and $\beta\geq 1$, $\zeta_0<0$. This is consistent with our expectations, as negative values of $\zeta_0$ ensure that the sign of the derivative of $g_{00}$ remains consistent in both the interior and exterior metrics, which is a requirement for the matching to be smooth.

We can now analyze the first junction condition in Eq. \eqref{junction}, i.e., $\left[h_{\alpha\beta}\right]=0$. For that, start by noticing that the angular parts of the metrics in Eqs.\eqref{metrici} and \eqref{metrice} are the same, therefore the angular components of the induced metric $h_{\alpha\beta}$ are trivially continuous, thus one just needs to analyze the $h_{00}$ components independently. Let us write the condition $\left[h_{00}\right]=0$ explicitly, which takes the following form 
\begin{equation}\label{cond3}
C e^{\zeta_0\left(\frac{r_0}{r}\right)^\alpha}=\left(1-\frac{2M}{r}\right).
\end{equation}
From the second junction condition, we have obtained the expressions for $r_\Sigma$ and $\zeta_0$, given by Eqs.\eqref{rsigma} and \eqref{zsigma}, respectively. These expressions are now introduced in Eq.\eqref{cond3}, in order to solve with respect to the constant $C$. This way, one obtains
\begin{equation}\label{csigma}
C=\left[1-\left(\frac{2M}{r_0}\right)^{1+\frac{1}{\beta}}\right]e^{-\alpha\left[\left(\frac{r_0}{2M}\right)^{1+\frac{1}{\beta}}-1\right]}.
\end{equation}
Once again, from the condition $r_0>2M$, we conclude that the constant $C$ is always strictly positive, independently of the values of $\alpha\geq1$ and $\beta\geq1$, which preserves the correct metric signature.  

Finally, let us summarize the obtained results. For a given choice of $r_0>2M$, $\alpha\geq 1$, and $\beta\geq 1$, the second junction condition $\left[K_{\alpha\beta}\right]=0$ gives the radius $r_\Sigma$ at which the matching must be performed, Eq.\eqref{rsigma}, as well as the value for the $\zeta_0$, Eq.\eqref{zsigma}). Using these expressions in the first junction condition $\left[h_{\alpha\beta}\right]$, one obtains the value of the constant $C$, Eq.\eqref{csigma}, which allows for the complete spacetime metric to be continuous at $\Sigma$.

%%%%%%%%%%%%%%%%%%%%%%%%%%%%%%%%%%%%%%%%%%%%%%%%%%%%%%
\section{Extensions to higher powers of $\mathcal T$}\label{sec:extensions}
%%%%%%%%%%%%%%%%%%%%%%%%%%%%%%%%%%%%%%%%%%%%%%%%%%%%%%

In the previous section \ref{sec:matching}, we explored wormhole solutions within the framework of the $f\left(R,\mathcal T\right)$ theory, assuming linearity in both $R$ and $\mathcal T$ as given by Eq.\eqref{linear_f}. Using a recursive method, we solved the systems of equations \eqref{eqrho}--\eqref{eqpt} to obtain wormhole solutions that satisfy the energy conditions, Eqs.\eqref{NEC}--\eqref{DEC}. We also performed a matching between the interior wormhole solution and an exterior vacuum spacetime to preserve the locality of these solutions. Now, in this section, our aim is to demonstrate that if the form of the function $f\left(R,\mathcal T\right)$ remains separable and linear in $R$, with an additional term proportional to higher powers of $\mathcal T$, the previous analysis can be straightforwardly generalized to this case. Then, the previous approach is still applicable for finding wormhole solutions in this case, with the only drawback being a longer computational time.

Let us start this analysis by considering the following expression for the function $f\left(R,\mathcal T\right)$
\begin{equation}\label{eq:frtnew}
    f\left(R,\mathcal T\right)=R+\gamma \mathcal T + \sigma \mathcal T^n,
\end{equation}
from which the field equations and the conservation equation given in Eqs.\eqref{geo_field} and  \eqref{geo_conservation}, respectively, take the following forms
\begin{eqnarray}\label{eq:new_field}
    G_{ab}=8\pi T_{ab}-\gamma\left(\Theta_{ab}-\frac{1}{2}g_{ab}\mathcal T\right)- \sigma \mathcal{T}^{n-1}\left(n\Theta_{ab}-\frac{1}{2}g_{ab}\mathcal T\right),
\end{eqnarray}
\begin{eqnarray}
    8\pi \nabla_b T^{ab}=\gamma\nabla_b\left(\Theta^{ab}-\frac{1}{2}g^{ab}\mathcal T\right)+\sigma\nabla_b\left(n\mathcal T^{n-1}\Theta^{ab}-\frac{1}{2}g^{ab}\mathcal T^n\right),
\end{eqnarray}
respectively. 

Taking into account the same metric as previously, Eq.\eqref{metric}, and a matter distribution that has the same form as previously, Eq. \eqref{def_matter}, from Eq.\eqref{eq:new_field} we obtain the field equations that feature several additional terms in comparison to Eqs.\eqref{eqrho}--\eqref{eqpt}, resulting in extremely lengthy equations that we do not write explicitly here. Nevertheless, in the next subsections we outline the fundamental differences for this case. 

\subsection{Wormhole solutions with extensions to higher powers of $\mathcal{T}$}

For a linear form of the function $f\left(R,\mathcal T\right)$ as employed in Sec.\ref{sec:theory}, the derived field equations \eqref{eqrho}--\eqref{eqpt} involve, at most, quadratic terms in the matter variables $\rho$, $p_r$, and $p_t$, thus resulting in a set of, at most, eight possibly complex solutions for these quantities. Now, when we consider higher powers of $\mathcal T$, we consequently obtain higher powers of the matter quantities in the field equations, which then results in a larger set of solutions. However, note that since the function $f\left(R,\mathcal T\right)$ does not feature any crossed terms in $R$ and $\mathcal T$, the resultant relationship between the matter quantities is still algebraic. Otherwise, we would find a differential relation if crossed terms were present, due to the fact that third term on the left hand side of Eq.\eqref{geo_field} would not vanish. Therefore, the recursive method presented in subsection \ref{subsec:worm sol} used to obtain wormhole solutions is still applicable to this case.

\subsection{Junction conditions}

In Chapter \ref{ch:results}, we present the solutions obtained by considering a function $f\left(R,\mathcal T\right)$ that is quadratic in $\mathcal T$. Once again, the matter components do not vanish across the entire range of the radial coordinate, indicating that these solutions are not localized. Consequently, these solutions need to be matched with an external spacetime corresponding to a vacuum. To achieve this, we first need to analyze the implications of having a higher-order power in $\mathcal T$ in the function $f\left(R,\mathcal T\right)$, to derive the junction conditions for this case.

When considering a higher-order power of $\mathcal T$ in the action, the modified field equations are given in Eq. \eqref{eq:new_field}. Note that these equations present an additional term proportional to $\sigma$, in comparison to their linear counterparts Eqs.\eqref{field}. Moreover, this additional term possesses products such as $\mathcal T^{n-1}\Theta_{ab}$ and powers of $\mathcal T^n$. Regarding the junction conditions, remember that in section \ref{sec:matching}, we have shown that $\mathcal{T}$ is only well-defined in the distributional formalism if the matching is smooth, i.e., in the absence of a thin-shell at the hypersurface $\Sigma$, leading to $S_{ab}=0$. Therefore, this restriction implies that $\mathcal T$ as well as the  auxiliary tensor $\Theta_{ab}$ given in Eq. \eqref{def_theta_2}, are both completely regular, meaning that they only feature terms proportional to $\theta\left(l\right)$, and terms such as $\delta\left(l\right)$ are absent from their expressions. As a result, the products between $\mathcal T$ and $\Theta_{ab}$, as well as the powers $\mathcal T^n$, also preserve this same regularity. This means that, no additional junction conditions arise from the addition of a higher-order power-law of $\mathcal T$ into the function $f\left(R,\mathcal T\right)$. Therefore, we have the same two junction conditions for this case, Eq.\eqref{junction}.

%%%%%%%%%%%%%%%%%%%%%%%%%%%%%%%%%%%%%%%%%%%%%%%%%%%%%%
\chapter{Results}\label{ch:results}
%%%%%%%%%%%%%%%%%%%%%%%%%%%%%%%%%%%%%%%%%%%%%%%%%%%%%%

%%%%%%%%%%%%%%%%%%%%%%%%%%%%%%%%%%%%%%%%%%%%%%%%%%%%%%
\section{Linear $f(R,\mathcal{T})$ gravity theory}
%%%%%%%%%%%%%%%%%%%%%%%%%%%%%%%%%%%%%%%%%%%%%%%%%%%%%%

\subsection{Wormholes solutions}

Let us start this chapter by presenting specific examples of the obtained solutions for the linear $f(R,\mathcal{T})$ gravity. Regarding the linear $f\left(R,T\right)$ gravity theory \cite{Rosa:2022osy}, it was found that the solutions that satisfy all of the energy conditions exist for negative values of the coupling constant $\gamma$. In our case, we obtain the same result, with the distinction that our solutions, despite being asymptotically flat, do not approach vacuum asymptotically. 

To illustrate this clearly, let us examine a specific set of parameters: $\alpha=\beta=-\gamma=1$, $r_0=3M$, and $\zeta_0=-6/5$.\footnote{While the selection of $\zeta_0=-6/5$ at this stage might appear arbitrary, we chose this particular value for reasons that become evident in the next subsection. For the current discussion, a broad range of other values for $\zeta_0$ including positive ones, would result in a qualitatively similar solution.} In Fig. \ref{fig:solution}, we plot the matter quantity $\rho$ and the combinations $ \rho+p_r$, $\rho+p_t$, $\rho+p_r+2p_t$, $\rho -|p_r|$ and $\rho-|p_t|$ of the solution that results from this particular combination that satisfies all of the energy conditions previously mentioned.

Additionally, to illustrate the influence of the parameters $\alpha$ and $\beta$ on the solutions, in Fig. \ref{fig:solutionab} we plot the quantities $\rho$, $p_r$, and $p_t$ for the set of parameters $\gamma=-1$, $r_0=3M$, and $\zeta_0=-6/5$ under different combinations of $\alpha$ and $\beta$.

\begin{figure}
  \centering
  \begin{subfigure}[b]{0.48\textwidth}
    \includegraphics[width=\textwidth]{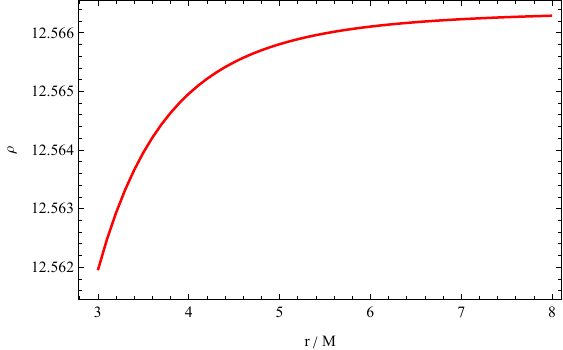}
    \caption{$\rho$}
  \end{subfigure}
  \begin{subfigure}[b]{0.48\textwidth}
    \includegraphics[width=\textwidth]{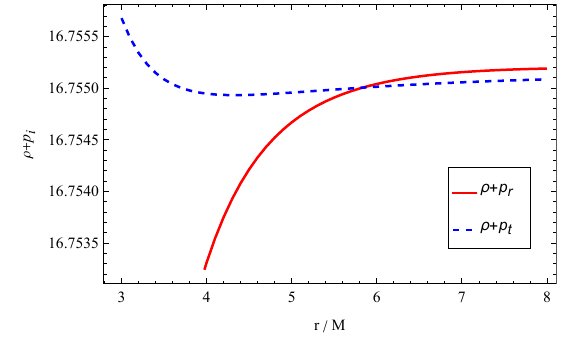}
    \caption{$\rho+p_i$, with $i=r,t$}
  \end{subfigure}
  
  \begin{subfigure}[b]{0.48\textwidth}
    \includegraphics[width=\textwidth]{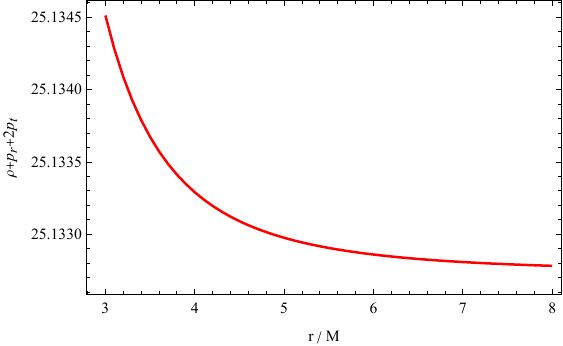}
    \caption{$\rho+p_r+2p_t$}
  \end{subfigure}
  \begin{subfigure}[b]{0.48\textwidth}
    \includegraphics[width=\textwidth]{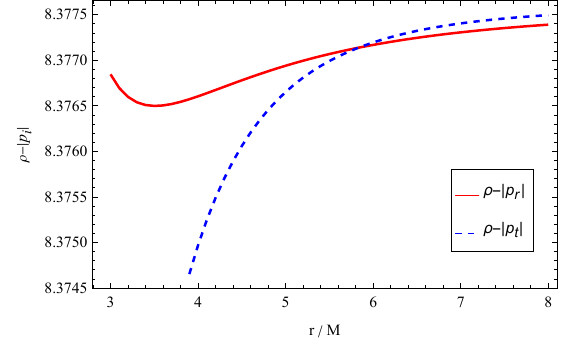}
    \caption{$\rho-|p_i|$, with $i=r,t$}
  \end{subfigure}
  \caption{Matter density $\rho$ and the combinations of energy conditions, namely $\rho+p_r$, $\rho+p_t$, $\rho+p_r+2p_t$, $\rho-|p_r|$ and $\rho-|p_t|$, are presented for the set of parameters $\alpha=\beta=-\gamma=1$, $r_0=3M$ and $\zeta_0=-\frac{6}{5}$.}
  \label{fig:solution}
\end{figure}

\begin{figure*}
  \centering
  \includegraphics[width=0.32\textwidth]{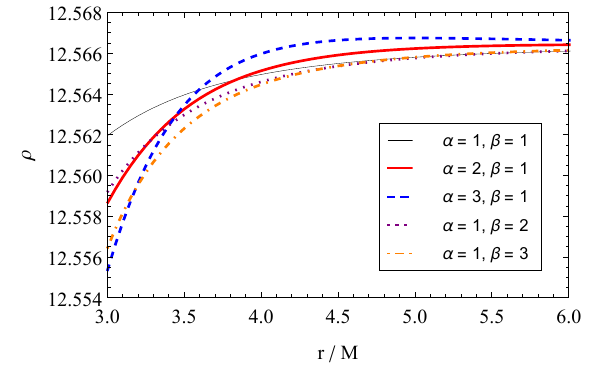}
  \hspace{0.001\textwidth} % Adjust the horizontal space between figures
  \includegraphics[width=0.32\textwidth]{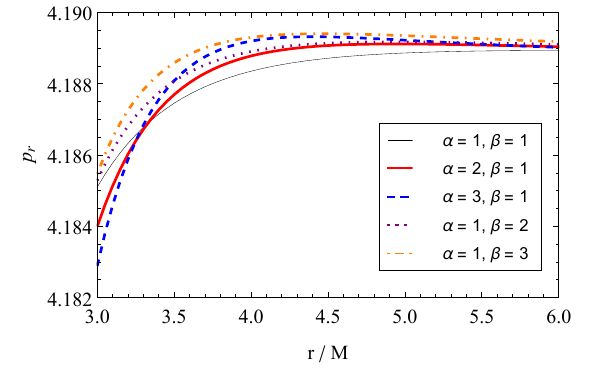}
  \hspace{0.001\textwidth} % Adjust the horizontal space between figures
  \includegraphics[width=0.32\textwidth]{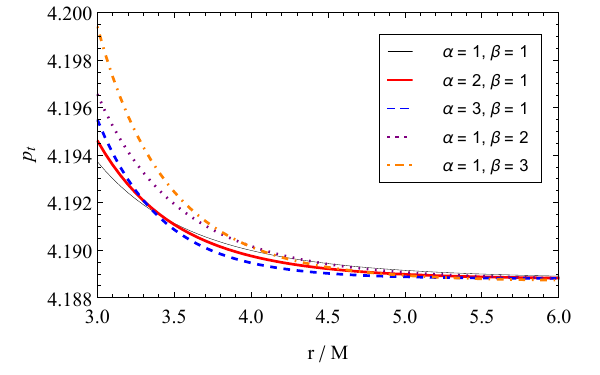}
  \caption{Matter quantities $\rho$, $p_r$, and $p_t$ are presented for the set of parameters $\gamma=1$, $r_0=3M$, and $\zeta_0=-\frac{6}{5}$ under different combinations of $\alpha$ and $\beta$.}
  \label{fig:solutionab}
\end{figure*}

Note that the solutions presented, while asymptotically flat, exhibit non-vanishing matter components throughout the entire radial coordinate range, meaning that these are not localized. This is precisely what motivated us to the derive the junction conditions for this theory, so one could match these solutions with an exterior vacuum spacetime at some finite radius, resulting then in localized wormhole solutions. 

\subsection{Junction conditions}\label{res:jc}

Let us explore a couple of specific applications of the junction conditions. Consider the case with parameters $r_0=3M$, $\alpha=1$, and $\beta=1$. For this particular set of parameters, Eq. \eqref{rsigma} determines the matching radius as $r_\Sigma=\frac{9}{2}M$, Eq. \eqref{zsigma} gives $\zeta_0=-\frac{6}{5}$, and Eq. \eqref{csigma} sets $C=\frac{5}{9}e^{\frac{4}{5}}$. The left panel of Fig. \ref{fig:matching} illustrates the $g_{00}$ component of the interior, exterior, and matched metrics. Alternatively, for a slightly different behavior, consider $\alpha=10$ while maintaining $\beta=1$ and $r_0=3M$. In this scenario, the matching radius remains $r_\Sigma=\frac{9}{2}M$, while $\zeta_0$ and $C$ take values $\zeta_0\sim -4.61320$ and $C\sim0.601826$. The right panel of Fig. \ref{fig:matching} illustrates this solution. Overall, it is evident that in both cases the $g_{00}$ component smoothly transitions from the interior to the exterior metric, ensuring the continuity of both the induced metric and the extrinsic curvature. 

Let us now examine the radial component of the metric, $g_{rr}$. We consider the same combination of parameters as before with $\alpha=\beta=1$, where the other parameters have been already computed, and another combination with $\alpha=1$ and $\beta=3$, resulting in $r_\Sigma \sim 3.43414 M$, $\zeta_0 \sim -1.59637$, and $C\sim 1.68432$. For both sets of parameters, the $g_{rr}$ components of the metric are shown in Fig. \eqref{fig:matchingR}. Although the radial component of the metric is not directly affected by the junction conditions, as both the induced metric $h_{\alpha\beta}$ and the extrinsic curvature $K_{\alpha\beta}$ are 3-dimensional tensors on the hypersurface $\Sigma$, one can observe that $g_{rr}$ is continuous but not differentiable at $r=r_\Sigma$. This continuity in $g_{rr}$ is expected when considering its dependence on the mass function inside a spherical hypersurface of radius $r$, denoted as $m\left(r\right)$, i.e., $g_{rr}=\left(1-\frac{2m\left(r\right)}{r}\right)^{-1}$. Using this, we find $m\left(r\right)=\frac{r_0}{2}\left(\frac{r_0}{r}\right)^\beta$ (see to Eqs.\eqref{metrici} and \eqref{zbfunctions}). In fact, since the matching between the interior and exterior spacetimes is smooth, thus without a thin-shell, the mass function $m\left(r\right)$ is expected to be continuous at $r_\Sigma$, ensuring the continuity of the $g_{rr}$ component of the metric.

\begin{figure*}
    \centering
    \begin{minipage}[b]{0.48\textwidth}
        \centering
        \includegraphics[scale=0.8]{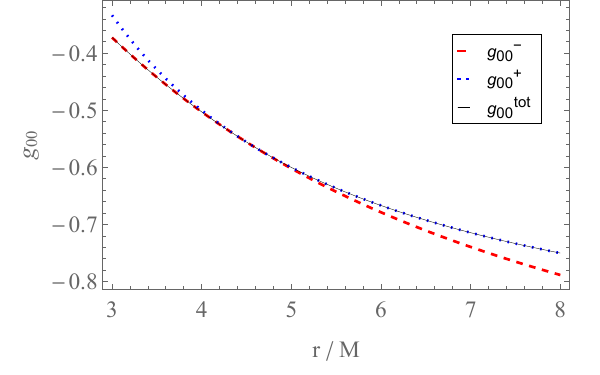}
    \end{minipage}
    \hfill
    \begin{minipage}[b]{0.48\textwidth}
        \centering
        \includegraphics[scale=0.8]{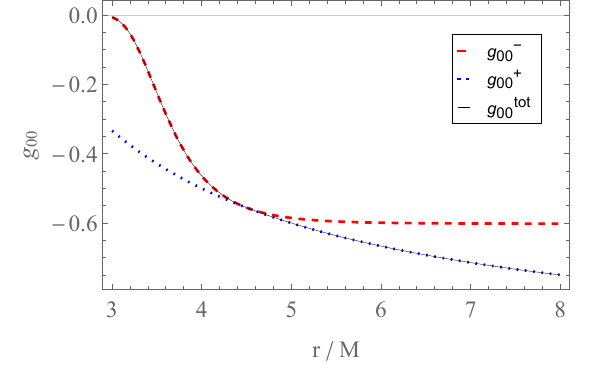}
    \end{minipage}
    \caption{The components $g_{00}$ of the interior wormhole spacetime given in Eq.\eqref{metrici} (red dashed curve) and the exterior Schwarzschild spacetime described by Eq.\eqref{metrice} (blue dotted curve) are shown for $\beta=1$, $r_0=3M$, and $\alpha=1$ (left panel) or $\alpha=10$ (right panel). The thin black line represents the solution $g_{00}^{\text{tot}}$ obtained through the matching between the interior and exterior solutions at $r=r_\Sigma$.}
    \label{fig:matching}
\end{figure*}

\begin{figure*}
    \centering
    \begin{minipage}[b]{0.48\textwidth}
        \centering
        \includegraphics[scale=0.8]{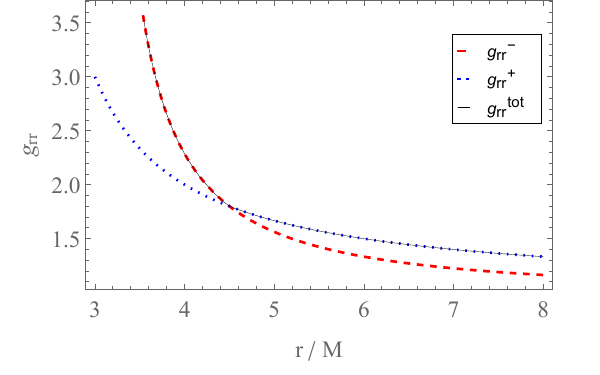}
    \end{minipage}
    \hfill
    \begin{minipage}[b]{0.48\textwidth}
        \centering
        \includegraphics[scale=0.8]{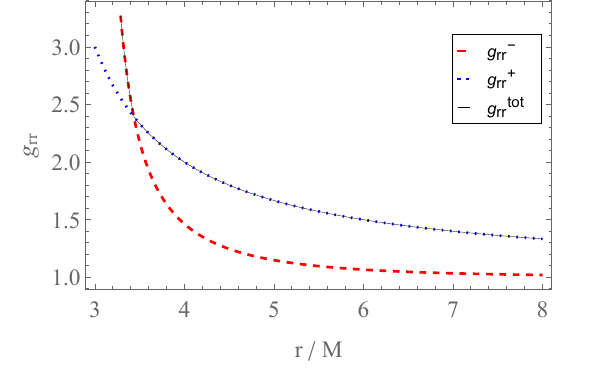}
    \end{minipage}
    \caption{The components $g_{rr}$ of the interior wormhole spacetime described by Eq.\eqref{metrici} (red dashed curve) and the exterior Schwarzschild spacetime given in Eq.\eqref{metrice} (blue dotted curve) are shown for $\alpha=1$, $r_0=3M$, and $\beta=1$ (left panel) or $\beta=3$ (right panel). The thin black line corresponds to the solution $g_{rr}^{\text{tot}}$ obtained through the matching between the interior and exterior solutions at $r=r_\Sigma$.}
    \label{fig:matchingR}
\end{figure*}

%%%%%%%%%%%%%%%%%%%%%%%%%%%%%%%%%%%%%%%%%%%%%%%%%%%%%%
\section{Extensions to higher powers of $\mathcal{T}$}
%%%%%%%%%%%%%%%%%%%%%%%%%%%%%%%%%%%%%%%%%%%%%%%%%%%%%%

\subsection{Wormholes solutions}

Similar to the findings in the linear version of the theory, one can observe that wormhole solutions satisfying all of the energy conditions also exist in theories described by functions $f\left(R,\mathcal T\right)$ with higher powers of $\mathcal T$. For instance, let us consider the same wormhole metric as before, characterized by the parameters $\alpha=\beta=1$, $r_0=3M$, and $\zeta_0=-6/5$. In this case, we choose a specific form of $f\left(R,\mathcal T\right)$ that is linear in $R$ and quadratic in $\mathcal T$, defined by the parameters $\gamma=0$, $\sigma=-1$, and $n=2$. The matter quantity $\rho$ and the combinations of energy conditions, namely $\rho+p_r$, $\rho+p_t$, $\rho+p_r+2p_t$, $\rho-|p_r|$, and $\rho-|p_t|$, are presented in Fig. \ref{fig:solution2}. As these solutions also lack localization and require matching with an external vacuum, we proceed to analyze the junction conditions for this case in the next subsection.

\begin{figure}
  \centering
  \begin{subfigure}[b]{0.48\textwidth}
    \includegraphics[width=\textwidth]{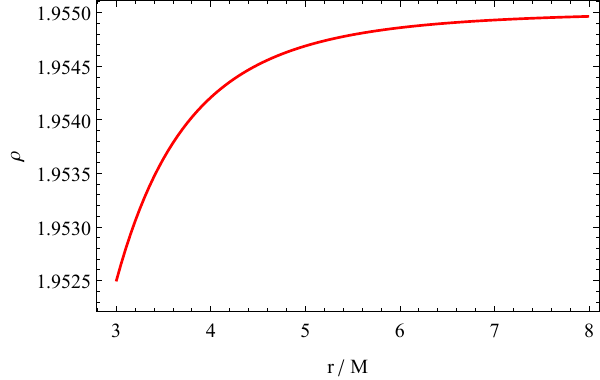}
    \caption{$\rho$}
  \end{subfigure}
  \begin{subfigure}[b]{0.48\textwidth}
    \includegraphics[width=\textwidth]{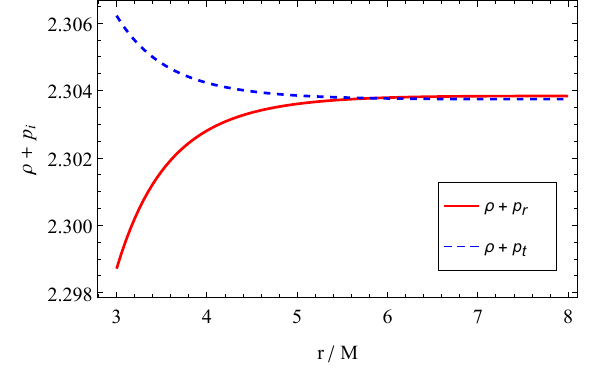}
    \caption{$\rho+p_i$, with $i=r,t$}
  \end{subfigure}
  
  \begin{subfigure}[b]{0.48\textwidth}
    \includegraphics[width=\textwidth]{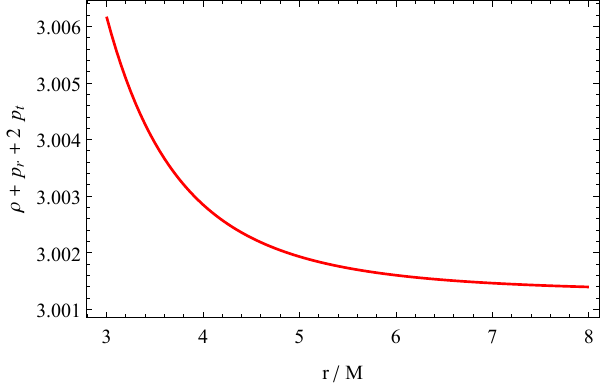}
    \caption{$\rho+p_r+2p_t$}
  \end{subfigure}
  \begin{subfigure}[b]{0.48\textwidth}
    \includegraphics[width=\textwidth]{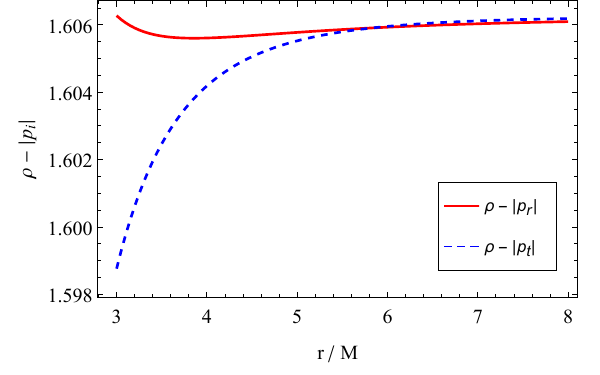}
    \caption{$\rho-|p_i|$, with $i=r,t$}
  \end{subfigure}
  \caption{Matter density $\rho$ and combinations of energy conditions, namely $\rho+p_r$, $\rho+p_t$, $\rho+p_r+2p_t$, $\rho-|p_r|$, and $\rho-|p_t|$, for the set of parameters $\alpha=\beta=-\sigma=1$, $\gamma=0$, $r_0=3M$, and $\zeta_0=-\frac{6}{5}$}
  \label{fig:solution2}
\end{figure}

Additionally, it is noteworthy that the inclusion of higher powers of $\mathcal T$ relaxes the constraints on the lower powers in the form of $f\left(R,\mathcal T\right)$. Remember that for the case of a linear form of this function, achieved with $\sigma=0$, in section \ref{subsec:worm sol} we concluded that solutions satisfying the energy conditions throughout the entire spacetime could only be obtained by choosing negative values for $\gamma$, as seen in Fig. \ref{fig:solution}. However, when a higher power of $\mathcal T$ is present, solutions meeting the energy conditions can be achieved even for positive values of $\gamma$. In Fig. \ref{fig:contours}, we present the values of matter quantities $\rho$, $p_r$, and $p_t$, as well as combinations of energy conditions, namely $\rho+p_r$, $\rho+p_t$, $\rho+p_r+2p_t$, $\rho-|p_r|$, and $\rho-|p_t|$, at the throat $r=r_0=3M$. We consider a function $f\left(R,\mathcal T\right)$ featuring both linear and quadratic terms in $\mathcal T$, i.e., $n=2$, for the same wormhole solutions with $\alpha=\beta=1$ and $\zeta_0=-6/5$. The plot illustrates the impact of varying values of $\gamma$ and $\sigma$. Note that, while both $\gamma$ and $\sigma$ influence the matter quantities at the throat, as long as $\sigma$ remains negative, $\gamma$ can take both positive and negative values, and the solution satisfies all energy conditions at the throat. Moreover, these solutions satisfying the energy conditions at the throat also satisfy these conditions for any radius beyond the throat, as demonstrated in Fig. \ref{fig:solution2} as a specific example.

\begin{figure*}
  \centering

  \begin{subfigure}[b]{0.32\textwidth}
    \includegraphics[width=\textwidth]{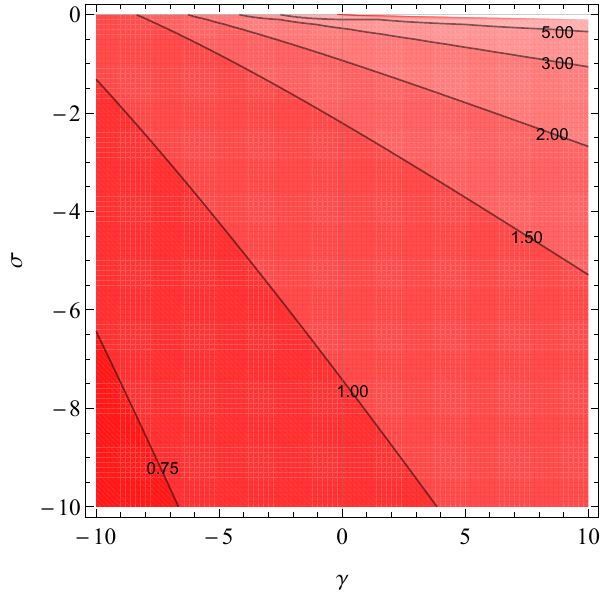}
    \caption{$\rho$}
  \end{subfigure}
  \begin{subfigure}[b]{0.32\textwidth}
    \includegraphics[width=\textwidth]{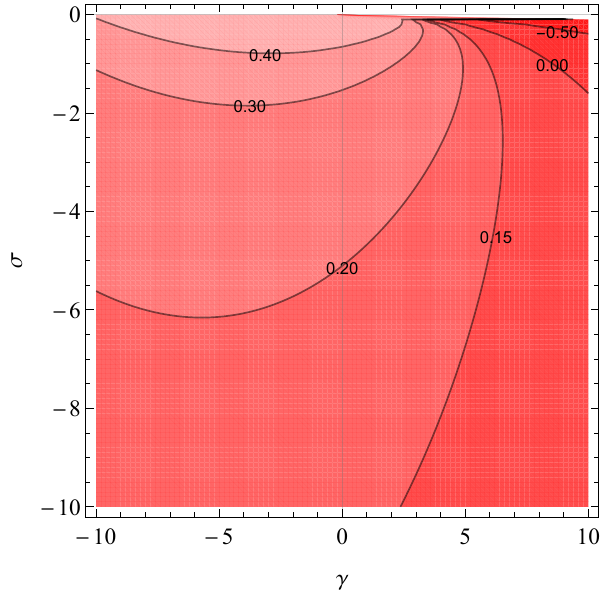}
    \caption{$p_r$}
  \end{subfigure}
  \begin{subfigure}[b]{0.32\textwidth}
    \includegraphics[width=\textwidth]{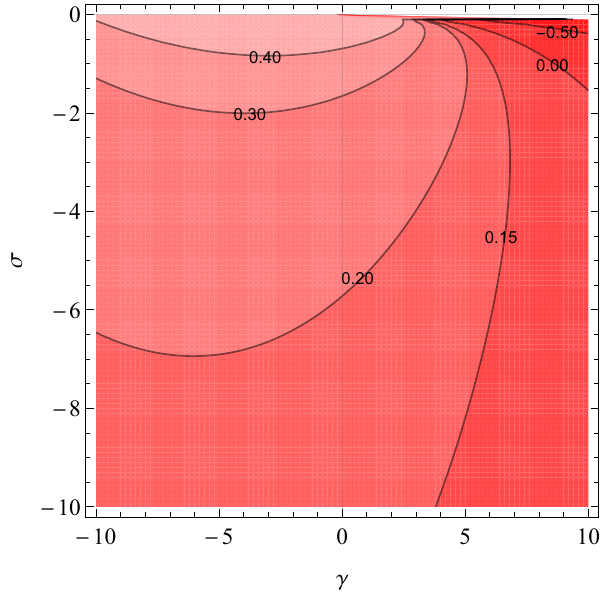}
    \caption{$p_t$}
  \end{subfigure}

  \begin{subfigure}[b]{0.32\textwidth}
    \includegraphics[width=\textwidth]{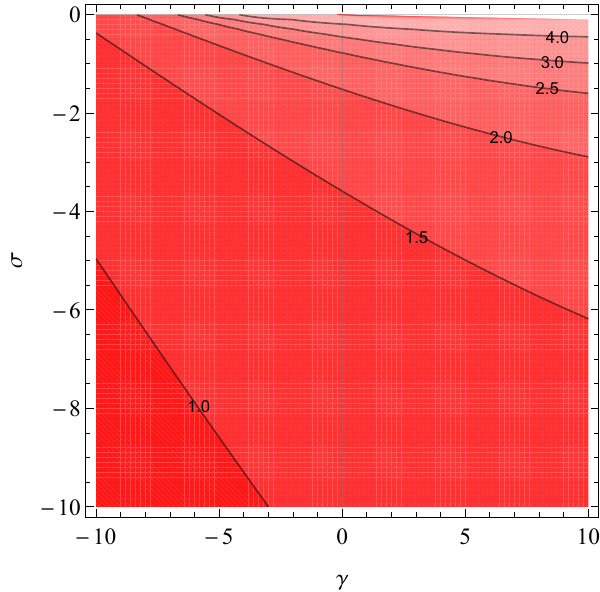}
    \caption{$\rho+p_r$}
  \end{subfigure}
  \begin{subfigure}[b]{0.32\textwidth}
    \includegraphics[width=\textwidth]{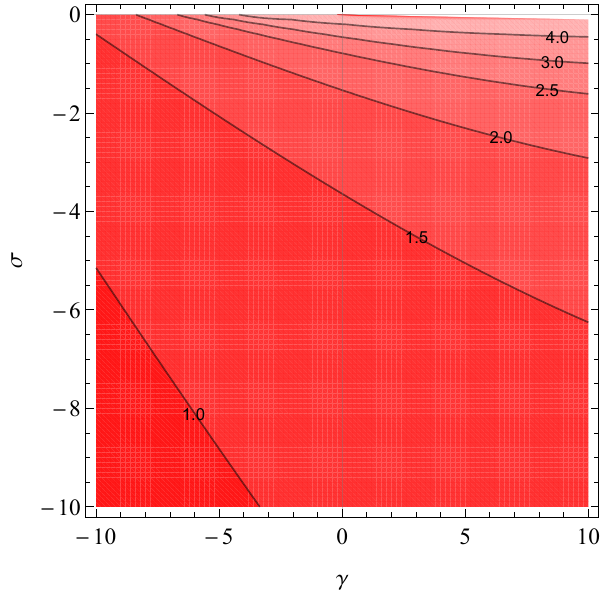}
    \caption{$\rho+p_t$}
  \end{subfigure}
  \begin{subfigure}[b]{0.32\textwidth}
    \includegraphics[width=\textwidth]{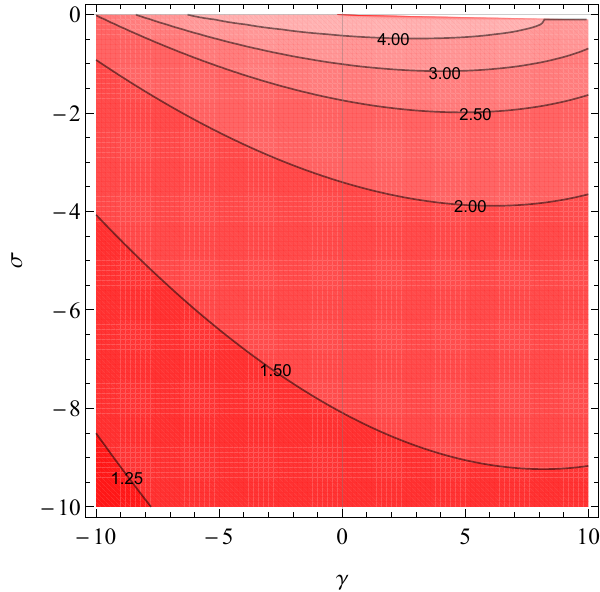}
    \caption{$\rho+p_r+2p_t$}
  \end{subfigure}

  \begin{subfigure}[b]{0.32\textwidth}
    \includegraphics[width=\textwidth]{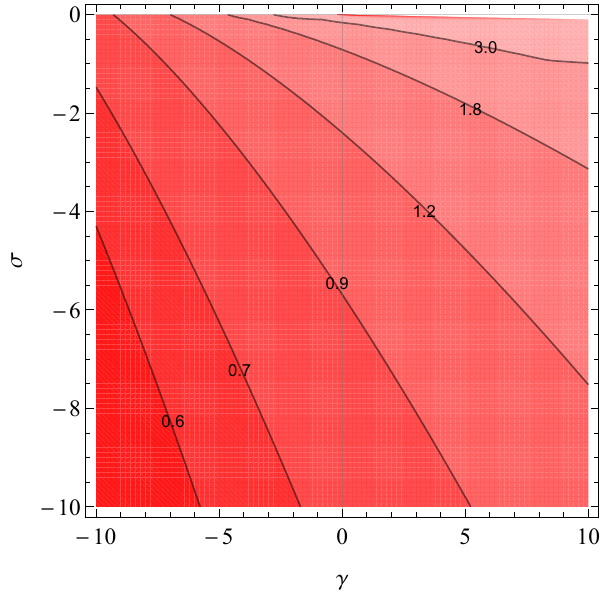}
    \caption{$\rho-|p_r|$}
  \end{subfigure}
  \begin{subfigure}[b]{0.32\textwidth}
    \includegraphics[width=\textwidth]{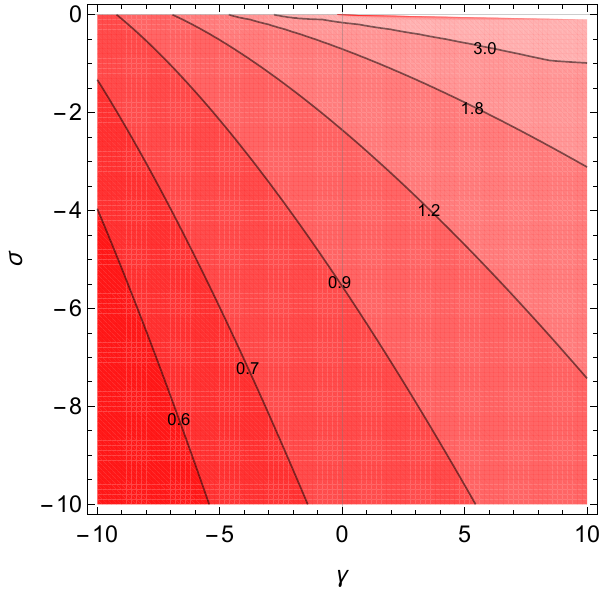}
    \caption{$\rho-|p_t|$}
  \end{subfigure}

  \caption{Values of the matter quantities $\rho$, $p_r$, and $p_t$, along with combinations of energy conditions, namely $\rho+p_r$, $\rho+p_t$, $\rho+p_r+2p_t$, $\rho-|p_r|$, and $\rho-|p_t|$, are evaluated at the throat $r_0$ for the parameters $\alpha=\beta=1$, $r_0=3M$, and $n=2$. Different values of $\gamma$ and $\sigma$ are considered to demonstrate the impact on these quantities.}
  \label{fig:contours}
\end{figure*}

\subsection{Junction conditions}

In the preceding subsection, we have derived wormhole solutions that satisfy the energy conditions throughout the entire spacetime, assuming a quadratic form for the function $f\left(R,\mathcal T\right)$. Once again, similar to the solutions obtained with a linear function, these solutions are not localized and necessitate matching with an external vacuum. However, remember that in section \ref{sec:extensions} we established identical junction conditions for this case as for the linear one. Consequently, the results obtained in subsection \ref{res:jc} are applicable here as well.

%%%%%%%%%%%%%%%%%%%%%%%%%%%%%%%%%%%%%%%%%%%%%%%%%%%%%%
\chapter{Conclusions}\label{ch:conclusion}
%%%%%%%%%%%%%%%%%%%%%%%%%%%%%%%%%%%%%%%%%%%%%%%%%%%%%%

In this thesis, we investigated traversable wormhole spacetimes within the framework of $f\left(R,\mathcal T\right)$ gravity, considering a linear model for both $R$ and $\mathcal T$. Our analysis revealed the existence of numerous traversable wormhole solutions that satisfy all energy conditions, from which we conclude that these solutions have a strong physical relevance. Given that the modified field equations exhibit quadratic dependence on the matter quantities $\rho$, $p_r$, and $p_t$, the theory yields eight independent potentially complex solutions for these quantities, some of which may have limited physical relevance. For that reason, to extract non-exotic wormhole solutions and understand the behavior of the matter quantities, we employed an iterative recursive algorithm.

A notable characteristic of the solutions obtained in this study is that, despite the spacetime metrics being asymptotically flat, the $f\left(R,\mathcal T\right)$ theory permits matter distributions that do not approach vacuum asymptotically, making them non-localized. However, through the application of the junction conditions in the theory, it becomes possible to localize these solutions. The derived junction conditions demonstrate that only a smooth matching is allowed in this theory, as the scalar $\mathcal T$ becomes singular in the presence of a thin-shell. In the case of a smooth matching, the junction conditions reduce to those of GR, which state the continuity of the induced metric and extrinsic curvature at the hypersurface separating the interior and exterior spacetime regions. Through this matching process, we have successfully obtained localized wormhole solutions that satisfy all energy conditions throughout the entire spacetime, enhancing their particular astrophysical significance.

The techniques presented in this study can be easily extended to more complicated dependencies of the function $f\left(R,\mathcal T\right)$ on $\mathcal T$, as long as there are no crossed terms between $R$ and $\mathcal T$. Since the relationship between the matter quantities $\rho$, $p_r$, and $p_t$ remains algebraic, it allows for the derivation of physically plausible wormhole solutions using the same recursive method. Additionally, the absence of crossed terms ensures that no extra junction conditions arise, facilitating the localization of solutions in a manner similar to the linear $\mathcal T$ counterpart, since the matching remains as a smooth one. To illustrate this, we applied these techniques to a quadratic version of the theory, successfully obtaining astrophysically relevant and localized wormhole solutions.

The $f\left(R,\mathcal T\right)$ gravity theory remains relatively unexplored in the context of wormhole physics, primarily due to the increased complexity of the field equations when considering more generalized forms of the function. Interesting extensions of this research could involve analyzing junction conditions for arbitrary forms of the function, particularly those including crossed terms of $R$ and $\mathcal T$. Such forms of the function would necessitate the development of new methods to solve the field equations, as they would involve differential relations between the matter fields. An alternative approach to explore more complex forms of the action involves considering the dynamically equivalent scalar-tensor representation of the theory. In this representation, the arbitrary dependence of the action on $R$ and $\mathcal T$ is replaced by two scalar fields, offering the advantage of reducing the order of the field equations to second-order. We plan to explore and address these issues in the future. 

In summary, our investigation has not only provided novel traversable wormhole solutions within the framework of the $f\left(R,\mathcal{T}\right)$ gravity theory but has also enriched the understanding of this theoretical framework. The theory not only offers physically plausible wormhole solutions but has also demonstrated success in addressing other compact objects \cite{Nari:2018aqs}-\cite{Sharif:2021uyc}, including black holes \cite{Chen:2019dip,Rudra:2020rhs}, and even in cosmological considerations \cite{Chamel:2016ynd}-\cite{Barbar:2019rfn} and \cite{Cipriano:2023yhv}. This success positions it as a promising avenue to modify General Relativity and now necessitates further exploration to thoroughly analyze its capabilities and limitations.

%
% Referências Bibliográficas
% 
%-----------------------------------------------------------------
%\renewcommand{\bibname}{Bibliografia}
%\bibliographystyle{babplain}      % basic style, author-year citations
%\bibliography{tese.bib}

\end{sloppy}
\end{document}